\documentclass[prb,notitlepage,superscriptaddress,showpacs,amsmath,amssymb,twocolumn]{revtex4-1}
\usepackage{hyperref,graphicx}
\usepackage{xcolor}
\usepackage{upgreek}
\usepackage{placeins}
\usepackage{siunitx}

\begin{document}

 \title{Velocity and confinement of edge plasmons in HgTe-based 2D topological insulators}

\author{Alexandre Gourmelon}
\altaffiliation{These authors contributed equally to this work}
\author{Elric Frigerio}
\altaffiliation{These authors contributed equally to this work}
\affiliation{Laboratoire de Physique de l'\'Ecole Normale Sup\'erieure, ENS, PSL Research University, CNRS, Sorbonne Universit\'e, Universit\'e de Paris, 24 rue Lhomond, 75231 Paris Cedex 05, France}
\author{Hiroshi Kamata}
\affiliation{Laboratoire de Physique de l'\'Ecole Normale Sup\'erieure, ENS, PSL Research University, CNRS, Sorbonne Universit\'e, Universit\'e de Paris, 24 rue Lhomond, 75231 Paris Cedex 05, France}
\affiliation{NTT Basic Research Laboratories, NTT Corporation, 3-1 Morinosato-Wakamiya, Atsugi, Kanagawa 243-0198, Japan}
\affiliation{JST, PRESTO, 4-1-8 Honcho, Kawaguchi, Saitama 332-0012, Japan}
\author{Lukas Lunczer}
\affiliation{Physikalisches Institut (EP3), Am Hubland, Universit\"at W\"urzburg, D-97074 W\"urzburg, Germany}
\affiliation{Institute for Topological Insulators, Am Hubland, Universit\"at W\"urzburg, D-97074 W\"urzburg, Germany}
\author{Anne Denis}
\author{Pascal Morfin}
\author{Michael Rosticher}
\author{Jean-Marc Berroir}
\author{Gwendal F\`eve}
\author{Bernard Pla\c cais}
\affiliation{Laboratoire de Physique de l'\'Ecole Normale Sup\'erieure, ENS, PSL Research University, CNRS, Sorbonne Universit\'e, Universit\'e de Paris, 24 rue Lhomond, 75231 Paris Cedex 05, France}
\author{Hartmut Buhmann}
\author{Laurens W. Molenkamp}
\affiliation{Physikalisches Institut (EP3), Am Hubland, Universit\"at W\"urzburg, D-97074 W\"urzburg, Germany}
\affiliation{Institute for Topological Insulators, Am Hubland, Universit\"at W\"urzburg, D-97074 W\"urzburg, Germany}
\author{Erwann Bocquillon}
\email{bocquillon@ph2.uni-koeln.de}
\affiliation{Laboratoire de Physique de l'\'Ecole Normale Sup\'erieure, ENS, PSL Research University, CNRS, Sorbonne Universit\'e, Universit\'e de Paris, 24 rue Lhomond, 75231 Paris Cedex 05, France}
\affiliation{II. Physikalisches Institut, Universit\"at zu K\"oln, Z\"ulpicher Str. 77, 50937 K\"oln, Germany}

\date{\today}

\begin{abstract}
High-frequency transport in the edge states of the quantum spin Hall (QSH) effect has to date rarely been explored, though it could cast light on the scattering mechanisms taking place therein. We here report on the measurement of the plasmon velocity in topological HgTe quantum wells both in the QSH and quantum Hall (QH) regimes, using harmonic GHz excitations and phase-resolved detection. We observe low plasmon velocities corresponding to large transverse widths, which we ascribe to the prominent influence of charge puddles forming in the vicinity of edge channels. Together with other recent works, it suggests that puddles play an essential role in the edge state physics and probably constitute a main hurdle on the way to clean and robust edge transport.
\end{abstract}


\maketitle

{\bf
}

Since its experimental discovery\cite{konig2007} in 2007, the quantum spin Hall (QSH) effect has been intensively studied, as its helical edge states offer an exciting playground for spin-polarized edge transport and topological superconductivity, with possible applications in both spintronics and topological quantum computation. Prominent transport signatures have been observed in HgTe quantum wells (QWs) such as non-local and spin-polarized transport \cite{roth2009, brune2012,calvo2017,bendias2018}, or the fractional Josephson effect in HgTe-based Josephson junctions \cite{bocquillon2016,deacon2017}. Alternatively, InAs/GaSb double QWs \cite{knez2011,akiho2016,irie2020} or layered materials such as bismuthene \cite{reis2017} or WTe$_2$ \cite{wu2018} have also been successfully identified as QSH insulators.

In this context, the investigation of high-frequency transport in 2D topological insulators, such as HgTe quantum wells here, is of high interest. The charge relaxation scales of the QSH edge carriers has been measured \cite{dartiailh2020} by microwave capacitance spectroscopy \cite{pallecchi2011,inhofer2017}, revealing that the edge states have a larger than predicted density of state, possibly due to neighboring puddles. It also suggests that the QSH effect could be enhanced in dynamical studies by exploiting the difference in transport or scattering timescales between topological and bulk carriers. We here explore another aspect, namely the velocity of plasmons propagating in the edge channels. In the quantum Hall effect of GaAs, InAs or graphene samples, the velocities of chiral edge magneto-plasmons have been widely studied, highlighting the role of intra- and inter-channel Coulomb interaction, of the confinement edge potential, or of the screening of Coulomb interaction by nearby metallic gates, and of dissipation in the bulk \cite{ashoori1992,zhitenev1993,sukhodub2004,gabelli2007,kamata2010,hashisaka2013,kumada2011,bocquillon2013,kumada2014a,freulon2015}. 

Here, we report on a systematic study of the velocities of plasmons in a HgTe quantum well, in the classical and quantum Hall regime (magnetic fields $B$ up to \SI{8}{\tesla}) of the conduction band, as well as in the topological gap where the quantum spin Hall effect takes place (at $B=0$). The measurements are performed in a dilution refrigerator at a temperature of $T\simeq\SI{20}{\milli\kelvin}$, for frequencies $f\simeq 3-\SI{10}{\giga\hertz}$. The (phase) velocity is accessed via the phase shift generated by the delay of a plasmon excitation propagating between a local source and a probe contact in the HgTe QWs. Though phase and group velocities may differ, they are known to coincide in the low-energy limit, which we experimentally confirm (see Supplementary Online Material). The phase shift can be rather accurately measured, even on small distances in which time-resolved techniques \cite{kumada2011,kumada2014,kamata2022} would be inoperable due to insufficient delay. This allows for a rather short propagation length $l$ (ranging between 3 and \SI{7}{\micro\meter}), in order to approach the ballistic length which does not exceed \SI{1}{\micro\meter} in our device. Finally, the electron density $n$ is tuned via to a gate voltage $V_g$ in a large range $n \simeq \num{0.5e11} - \SI{+5e11}{\per\square\centi\meter}$ in the conduction band. Our main observations can be summarized as follows. First, when the Fermi energy is in the conduction band and under the action of a perpendicular magnetic field $B$, we observe a transition in the magnetic-field dependent velocity, suggestive of the crossover between non-interacting edge states (Landauer-Büttiker picture, abbreviated as LB) and edge reconstruction under e-e interactions (Chklovskii-Shklovskii-Glazman regime\cite{chklovskii1992, armagnat2020}, denoted CSG). From the analysis of the velocity, we conclude that the edge states have a typical width $w_0$  of several microns at low fields, probably set by the electrostatic disorder, while the edge confinement itself occurs, as expected, on a typical scale $l\sim\SI{0.1}{\micro\meter}$ comparable to the distance $d$ between the gate and the HgTe layer. Second, we confirm this interpretation by an analysis of the observed low velocities in the QSH gap of the device.

The article is organized in three sections. In the first section, we introduce the device geometry, and the preliminary characterization of the samples via DC magneto-transport measurements. The microwave measurement setup is briefly described in the second section, together with the post-acquisition calibration process, with application to raw data. Finally, we detail several experimental results in the QSH and QH regime, and present a plausible interpretation based on the presence of charge puddles in the band gap of the material.

\section{Sample geometry and DC transport properties}

\begin{figure}[ht]
\centerline{\includegraphics[width=0.5\textwidth]{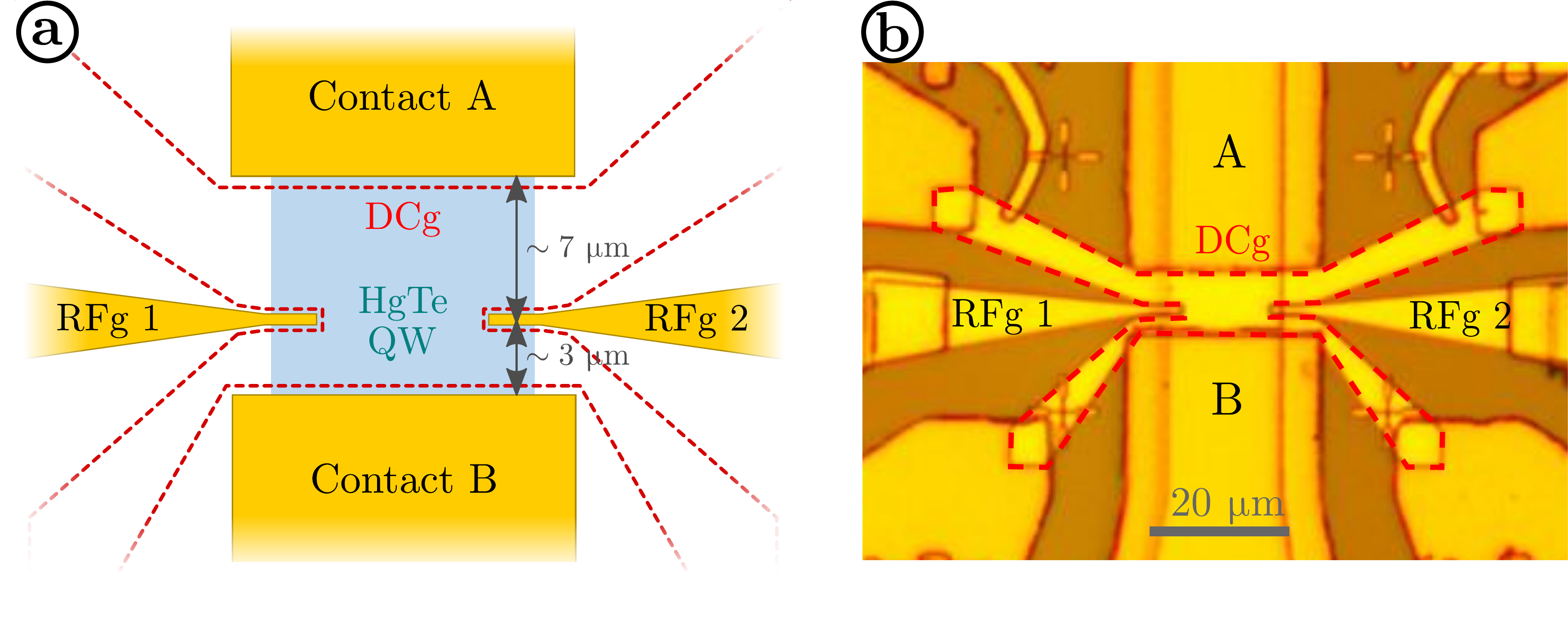}}
\caption{{\bf Sample geometry:} a) Sketch of the device: The light blue part is the HgTe mesa while the yellow parts (finger-gate electrodes RFg 1, 2 and ohmic contacts A, B) are made of gold. The red dashed region corresponds to the space covered by the top-gate DCg. A DC voltage $V_g$ is applied to RFg 1, 2 and DCg to uniformly tune the electron density $n$. b) Image taken with an optical microscope of the Au-gated sample, showing the different gates and contacts of the sample as sketched in a).
}\label{figure:sample}
\end{figure}

\paragraph{Samples --}
The samples are fabricated from HgTe/Cd$_{0.68}$Hg$_{0.32}$Te QWs grown by molecular beam epitaxy. The thickness $t$ of the QWs is \SI{8.5}{\nano\meter}. For such a thickness, the band structure consists of light electrons in the conduction band, and heavy holes in the valence band. A topological phase transition for thickness $t>t_c\simeq \SI{6.3}{\nano\meter}$ enforces the presence of QSH edge states in the gap of the QWs \cite{bernevig2006}. A gap of approx. \SI{26}{\milli\electronvolt} is predicted by ${\bf k}\cdot{\bf p}$ simulations of the band structure (estimated along the $k_x = \pm k_y$ direction in which it is minimal).
Additionally, the QW is protected by a Cd$_{0.68}$Hg$_{0.32}$Te capping layer of thickness \SI{50.5}{\nano\meter}. The QWs are first characterized using standard Hall-bar measurements, yielding a mobility of  $1-\SI{2e5}{\square\centi\meter\per\volt\per\second}$ (measured at a density $n\simeq2-\SI{3e11}{\per\square\centi\meter}$ in the conduction band). Three devices have been investigated and have given similar results. Each device comprises a rectangular mesa defined via a wet-etching technique \cite{bendias2018} to preserve the high crystalline quality and the high mobility of the epilayer. Low-resistance ohmic contacts are evaporated on either ends of the mesa, for both DC characterization and RF measurements. Two gold finger gates (denoted RFg 1 and RFg 2 in Fig.\ref{figure:sample}) are patterned with e-beam lithography, with a width $\delta\simeq \SI{800}{\nano\meter}$ and are used to locally and capacitively excite the underneath QW with high-frequency signals, while an additional gate for DC tuning of the electron density (DCg) covers the rest of the mesa. All gates are evaporated on top of a \SI{16}{\nano\meter}-thick HfO$_2$ insulating layer, grown by low-temperature atomic layer deposition (ALD)\cite{bendias2018}. The main text focuses on one device where the global gate DCg is made of Au (denoted "Au sample"), as presented in Fig.\ref{figure:sample}b. Another sample covered with a thin Pd global gate electrode (denoted "Pd sample") is also briefly discussed in the main text, with more data presented in the Supplementary Material. The results are very analogous, though our observations point towards larger puddles in the electrostatic landscape of the device.

\paragraph{Characterization of the transport properties --}
\begin{figure}[ht]
\centerline{\includegraphics[width=0.5\textwidth]{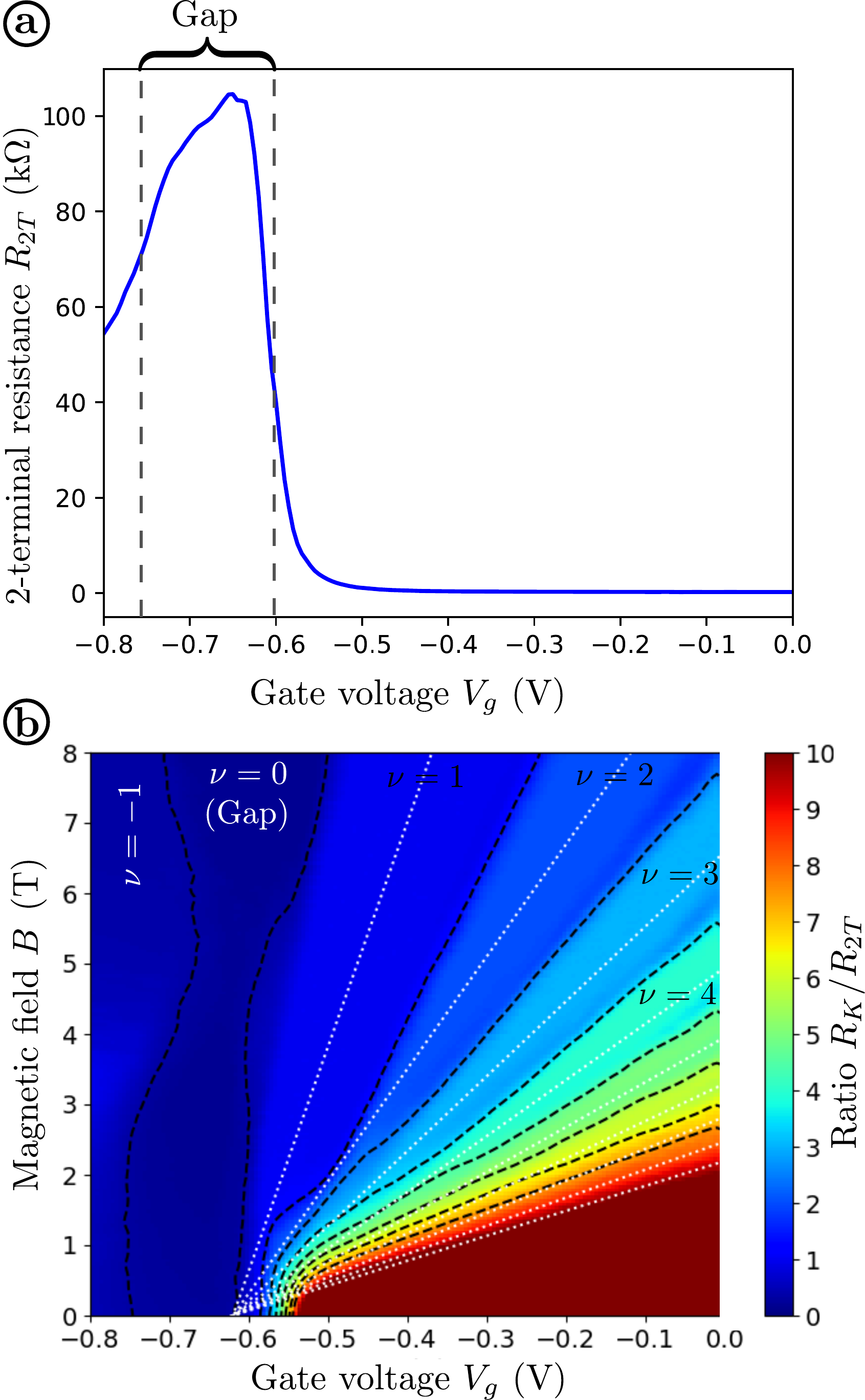}}
\caption{{\bf DC transport properties of the sample:} a) Two-terminal resistance $R_{2T}$ as a function of the gate voltage $V_g$, exhibiting a peak signaling the gap (indicated by the dashed lines), and the conduction and valence bands on either sides of this peak. b) 2D color map of the ratio $R_K/R_{2T}$ as a function of gate voltage $V_g$ and magnetic field $B$. The different filling factors $\nu$ are labelled, and the white dotted lines are the lines $B_\nu$ used to fit the carrier density $n$ as a function of gate voltage $V_g$ (see main text). The contours between the different QH plateaus are highlighted as dashed black lines. The color scale is intentionally saturated at a maximum value $R_K/R_{2T} = 10$, in order to distinguish more clearly the first QH plateaus.}
\label{figure:DCCharac}
\end{figure}
The two-terminal resistance $R_{2T}$ is measured at $B = \SI{0}{\tesla}$ as a function of the gate voltage $V_g$ (applied simultaneously to all three gates to preserve a uniform electron density), and presented in Fig. \ref{figure:DCCharac}a. It informs on the band structure and the position of the gap in the devices. Similarly to previous works, we identify the gap as a clear peak in $R_{2T}$, separating the conduction and valence band regimes. However the peak value of $R_{2T}$ is much larger than the expected quantized value $R_K/2 = h/2e^2\simeq \SI{12.9}{\kilo\ohm}$, and lies around \SI{120} {\kilo\ohm}. The full in-situ characterization of the mobility and electron density from the magnetoresistance of the devices is rendered impractical by the two-terminal geometry imposed by the microwave measurements. Indeed, the two-terminal resistance involves both the Hall and longitudinal resistance, which are easily separated in a four-terminal geometry. Nevertheless, one observes in Fig.\ref{figure:DCCharac}b that $R_{2T}$ clearly exhibits the quantum Hall plateaus. Assuming that these plateaus reach perfectly quantized values,  we write $R_{2T} = R_K/\nu + R_c$ at the center of each plateau of filling factor $\nu\in\mathbb{N}^*$, where $R_c$ is a contact resistance. While the contact resistance is estimated at $R_c\simeq\SI{100}{\ohm}$ in the conduction band, it appears to be much higher in the valence band (\SI{20}{\kilo\ohm}), presumably due to the formation of $p$-$n$ junctions near the $n$-doped contacts. The ratio $R_K/R_{2T}$ as a color plot in Fig.\ref{figure:DCCharac}b can then be used to fit the density by adjusting the set of lines $B_\nu= \frac{eR_K}{\nu}(\frac{c}{e} V_g + n_0)$ which define the exact integer filling factors. We then obtain the density $n(V_g)=\frac{c}{e} V_g + n_0$ where $c\simeq \SI{1}{\milli\farad\per\square\meter}$ is the gate capacitance per unit area (in agreement with theoretical estimate given the gate layer stack), and $n_0\simeq \SI{4.7e15}{\per\square\meter}$ the density at $V_g=0$.

\section{Microwave measurements and calibration}

\paragraph{Microwave setup --}

The phase shift in the device under test (DUT) is measured using a standard heterodyne detection method. An RF sine wave of frequency $f$ in the GHz regime is generated from an arbitrary waveform generator (AWG) and sent to the RFg of the sample through the microwave lines of the fridge. At the excitation finger gate RFg, the signal amplitude is typically \SI{1}{\milli\volt}. After being emitted by the finger gates RFg, the signal is collected by the two contacts (A and B). The signal is amplified by a cryogenic and room-temperature low-noise amplifiers before being sent to a heterodyne detection setup at room temperature. The signal is mixed with a local oscillator (LO) i.e. a sine wave generated signal generator detuned from the AWG output by 50 MHz. This mixing process converts the GHz signal coming from the sample to a 50 MHz signal which is then demodulated by a multi-channel fast acquisition card to obtain the {\sl in-phase} ($I$) and {\sl in-quadrature} ($Q$) parts of the signal $I\cos(2\pi ft)+Q\sin(2\pi ft)$, in each contact A and B. With this setup, it is possible to measure simultaneously signals at the two contacts A and B in the range of frequencies $f\simeq 3-\SI{10}{\giga\hertz}$ set by the cryogenic isolators placed before the cryogenic amplifiers. The full experimental setup is shown in the Supplementary Online Material for more detail.

\paragraph{Calibration of the raw data --}

\begin{figure}[h!]
\centerline{\includegraphics[width=0.45\textwidth]{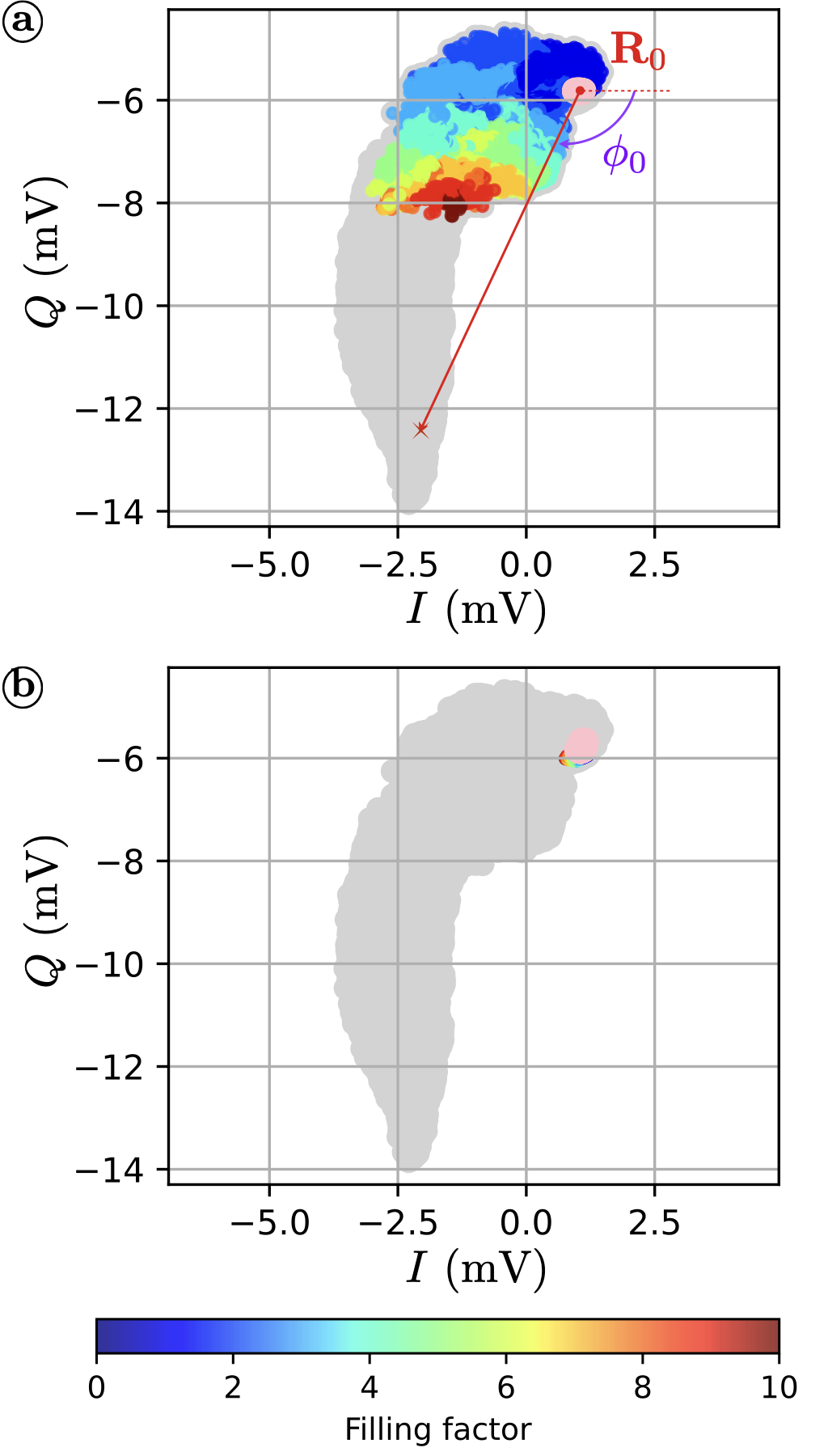}}
\caption{{\bf Calibration of the raw data:} In the Nyquist plane ($I$,$Q$), the same set of data points is represented in both panels, as light dots. a) Focusing on $B<0$, data points corresponding to filling factor $\nu=-1$ ($p$-type transport) are colored in pink, and for $\nu\in[0, 10]$ ($n$-type transport) colored according to the filling factor $\nu$ (as measured from the DC two-terminal resistance $R_{2T}$). Data points for $\nu>0$  occupy a large fraction of the total data set, indicating that the ac current flows from the finger gate RFg to the contact (A in this case) with a magnitude that depends on $\nu$. b) Focusing now on $B>0$, we use the same color coding for filling factors. The data points then occupy a very small fraction of the phase space. This indicates that the measured signal is independent of $\nu$ and is dominated by the stray coupling. Additionally, one can reverse the the polarity of the carriers and drive the sample into the $\nu=-1$ state (pink-colored data points). The parasite stray coupling ${\mathbf R}_0$ is indicated by a red dot. The red cross indicates the signal taken for $V_g = \SI{0}{\volt}$ and $B = 0$, which is used to determine the phase reference $\phi_0$.}
\label{figure:DataCalibration}
\end{figure}

The signal measured in the channels of the acquisition card need further calibration and reference: 1) The measured magnitude is offset by stray couplings on the chip and sample holder, which do not contain any physical information on the topological device. This parasitic contribution is measured in a situation where the DUT is known to be perfectly insulating, and then subtracted. 2) The phase is also affected by the propagation in the cables, and can not be directly used for computing the plasmon velocities. A phase reference need to be defined from a situation where currents propagate at a very high velocity in the DUT (much larger than the edge plasmon velocities). We describe in this paragraph how we proceed to these two steps, and how we control the validity of the underlying assumptions.

We first concentrate on the calibration of the amplitude and the subtraction of stray couplings. Such couplings are ubiquitous in microwave measurements, and can be as strong or even stronger than the physical signal through the DUT, in particular in high-impedances devices such as the ones considered here. Given the geometry of the device, reversing the magnetic field direction or the polarity of the carriers (from $n$- to $p$-regime) reverses the direction of the chiral edge states and thus nullifies the edge state signal measured in one of the two contacts. As an example, we take the situation of Fig.\ref{figure:DataCalibration}. There the data measured on contact $A$ at $f=\SI{4}{\giga\hertz}$ is shown in the Nyquist plane ($I$, $Q$) showing the in-phase and in-quadrature parts of the signal. For $B<0$, the current emitted by the finger gate RFg flows to contact A and then depends on the filling factor $\nu$. As a result, in Fig\ref{figure:DataCalibration}.a, the data points for $B<0$ span a wide zone (colored data points following the color bar). In contrast, for $B>0$, the data points of all filling factors are concentrated in a small area (see Fig.\ref{figure:DataCalibration}b), showing that no current flows from the RFg to contact $A$, and the data points then indicate the coordinate of the stray coupling in the $(I,Q)$ plane. Instead of reversing the field direction $B\to -B$, it is faster and equally accurate to reverse the polarity of the carriers and drive the sample into the $\nu=-1$ state (pink-colored data points in Fig.\ref{figure:DataCalibration}), allowing to subtract a reference vector ${\bf R}_0=(I_0,Q_0)$ indicated by the red dot in Fig.\ref{figure:DataCalibration}a. Thus we can measure and subtract the stray coupling with an estimated accuracy of a few \%.

We now explain how we reference the phase. Any signal passing through the DUT acquires a phase $\phi_0+\phi$  with $\phi$ inversely proportional to $v$. Thus, we define the phase reference $\phi_0$ in a situation where plasmons are considered infinitely fast ($\phi\to0$). To this end, we consider that 2D plasmons of the conduction band (in the absence of magnetic fields) propagate at a very large velocity (often reported $v_{\rm max}\gtrsim\SI{2e7}{\meter\per\second}$ in similar semi-conducting systems). In Fig.\ref{figure:DataCalibration}, this corresponds to a rotation of angle $\phi_0$ of the phase. Though the phase reference is rather roughly defined, it is precise enough for the study of all velocities $v\ll v_{\rm max}$. The velocities discussed later in this article validate a posteriori this approach, in line also with previous measurements in the QH regime of GaAs 2DEG \cite{kumada2011}.
 
\begin{figure*}[htbp!]
\centerline{\includegraphics[width=1\textwidth]{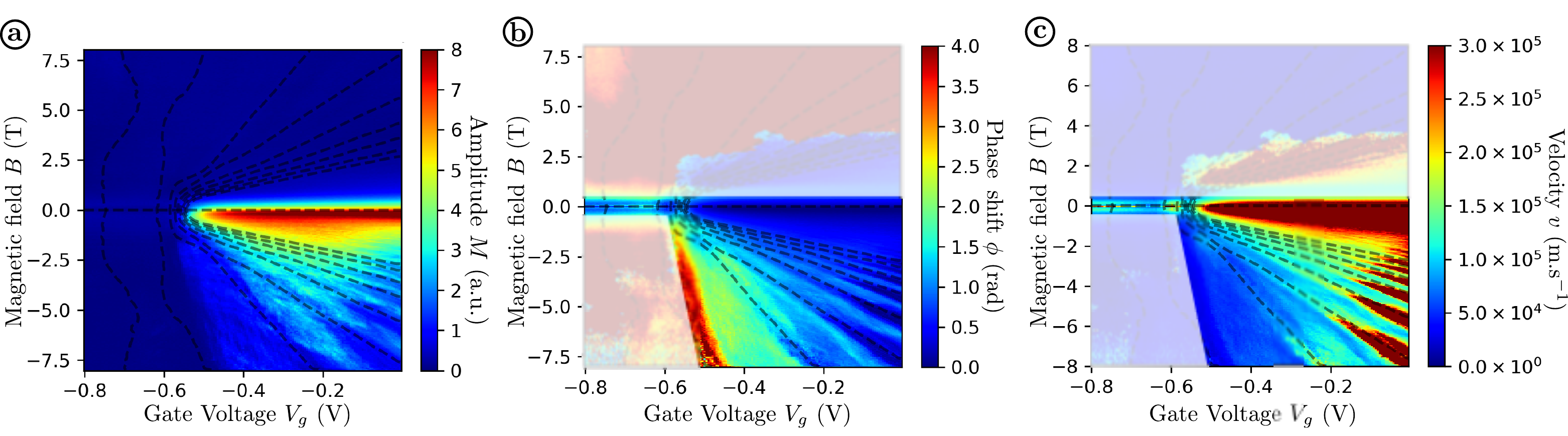}}
\caption{{\bf Calibrated amplitude, phase and velocity:}
Colormaps of the amplitude $M$ (a), phase $\phi$ (b) and velocity $v$ (c) as function of the gate voltage $V_g$ applied on DC$_g$ and the magnetic field $B$, obtained from the raw data presented in Fig. \ref{figure:DataCalibration}. The white shadings indicate regions where the signal amplitude $M$ is too small, so that $\phi$ and $v$ are not reliably computed.
}\label{figure:CalData}
\end{figure*}

These two calibration steps have been successfully conducted in numerous data sets. In such cases, the chirality of the QH edge channels manifests itself as a strong asymmetry with either the magnetic field directions, the choice of the contact (A or B) or the choice of the finger gate (RFg1 or RFg2) (see Supplementary Material for additional data and chirality maps). The phase also winds in a unique direction (clockwise). In Fig.\ref{figure:CalData}, we show the resulting calibrated amplitude $M$ of the microwave signal, its phase $\phi$, and the velocity $v$ calculated from the phase as $v=\frac{2\pi fL}{\phi}$, where $L$ is the propagation length between finger gate and contact. The amplitude is close to zero on one half of the plane (here for $B>0$), while it is strong on the other half (here for $B<0$), and gradually decays with increasing field. The different filling factors are clearly visible and agree well with those determined from the DC magnetoresistance. In the regions where the amplitude $M$ is sufficiently large, the phase can be unwrapped, allowing for the computation of the velocity in the same area.

However, some samples and data sets have resisted such an analysis, and exhibit asymmetric but not totally chiral behavior or do not allow to define an adequate phase reference. In agreement with our findings described later we attribute these phenomena to strong disorder in some samples, allowing for propagation of signal opposite to the expected propagation direction. We present problematic data sets in the Supplementary Material.

After calibrating the electron density $n$, the amplitude $M$ and phase $\phi $ of the microwave data, we now explore the variations of the velocity $v$ in the Hall regime as function of $n$, $B$ but also the filling factor $\nu=\frac{hn}{eB}$.

\section{Results -- Plasmon velocities}

In this section, we analyze the measured velocity, and discuss different interesting observations. When a perpendicular magnetic field is applied, a clear transition is observed between low and high-field regimes, which we attribute to the crossover between the LB non-interacting regime to the CSG regime\cite{chklovskii1992, armagnat2020} at high fields where e-e interactions are prominent. A careful study of both regimes then yields information on the role of puddles and edge confinement in the device, which is relevant for both the classical and quantum Hall regime, but also indicative of the physics of QSH edge states. Though the data is not as clear, we also confirm these observations in the gap of the quantum well at zero magnetic field, i.e. when QSH edge states dominate transport.

\paragraph{Plasmon confinement in the quantum Hall effect --}

\begin{figure*}[htbp!]
\centerline{\includegraphics[width=.8\textwidth]{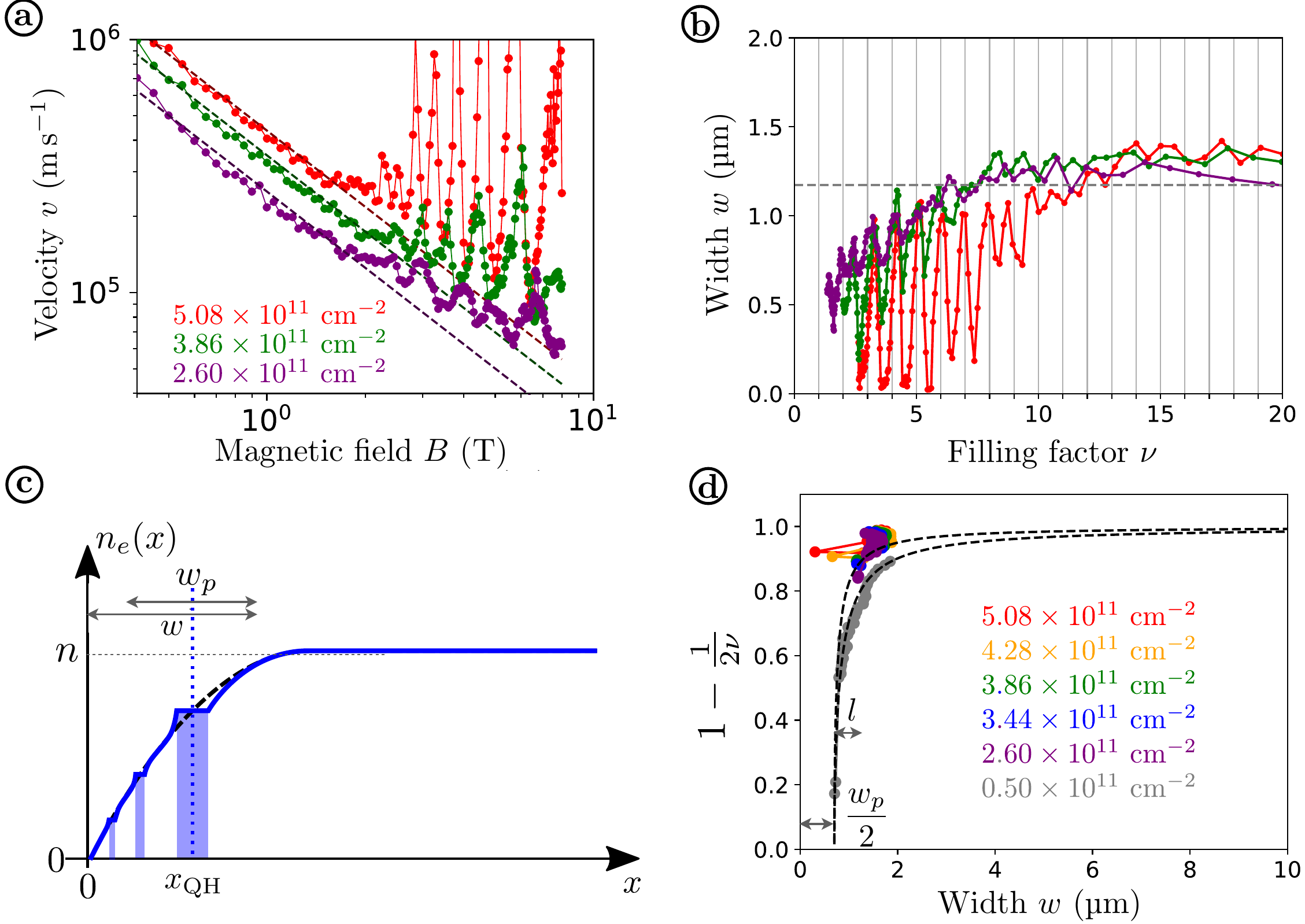}}
\caption{{\bf Velocity and plasmon transverse width in the Hall regime} a) Linecuts of the velocity $v$ as function of magnetic field $B$ for three values of the density $n$. The grey dashed line shows fits to the law $v\propto B^{-1}$, valid at low fields. In the high field region, $v$ exhibits strong oscillations, which become more pronounced as the density $n$ increases. b) Transverse width $w$ as a function of filling factor $\nu$. For large $\nu$, i.e. low fields, $w$ is approximately constant and independent of $n$. For low $\nu$, the width $w$ oscillates, showing minimum for integer filling factors $\nu\in\mathbb{Z}$. c) Sketch of the edge density profile $n_e(x)$, as a function of the distance $x$ from the edge: $n_e(x)$ saturates at $n_e(x)=n$ in the bulk of the material, and decreases to $n(x)=0$ at the edge. The blue shades indicate the compressible stripes while the white stripes are the incompressible ones. The bare plasmon width is given by the position of the innermost Landau level $x_{\rm QH}$, and is further increased by $w_p/2$ due to puddles. d) Normalized reconstructed edge profile $n_e(x)$ obtained by plotting $1-1/2\nu$ as a function of $w$ for all data triplets $(n, B, w)$. The obtained profiles are shown as colored dots for various values of the bulk density $n$. The dashed lines represents the heuristic edge profile $f(x)$ for two extreme admissible values of the depletion depth $l=60$ and \SI{150}{\nano\meter}.
}\label{figure:QHData}
\end{figure*}

We first turn to the study of plasmon velocities in the quantum Hall regime, i.e. when a perpendicular magnetic field is applied to the sample. In gated samples, the velocity of the edge magneto-plasmons can be simply written as
\begin{equation}
v=\frac{ned}{\epsilon Bw}=\frac{\sigma_{xy}}{C_{\rm QH}} \label{eq:velocity_QH}
\end{equation}
where $w$ is the transverse width of the edge plasmon, $\sigma_{xy}=ne/B$ the Hall conductance, and $C_{\rm QH}=\epsilon w/d$ the capacitance between gate and plasmon per unit length. This equation can be obtained from a microscopic derivation \cite{aleiner1994,johnson2003}. It also is a constitutive relation of a transmission line model for edge states \cite{burke2002}, connecting the line impedance $1/\sigma_{xy}$ and the velocity $v$ with the capacitance $C_{\rm QH}$.

Through Eq.(\ref{eq:velocity_QH}), the velocity $v$ provides insights into the confinement of plasmons on a width $w$ near the edges of the sample. In this context, the role of e-e interaction and screening in the progressive formation of edge states is well-understood since pioneering works in the 90s \cite{chklovskii1992, chamon1994}, and have recently been numerically revisited \cite{armagnat2020}.

In all measured devices, we observed two different behaviors depending on the strength of the magnetic field. At low field, the velocity is both proportional to $n$ and $1/B$ (as illustrated in Fig.\ref{figure:QHData}a for three different electron densities), in agreement with the Landauer-Büttiker model. In this model, a large number of edge states are uniquely defined by the edge confinement profile, while screening and reconstruction from e-e repulsion are irrelevant. This allows to define the $n$-independent width $w_0$ of the plasmon in this regime, and we find $w_0\simeq \SI{1.2}{\micro\meter}$ in the Au-gated sample ($w_0\simeq \SI{4.6}{\micro\meter}$ for the Pd-gated sample). 

This transverse width is much greater than the distance to the gate $d\simeq \SI{50}{\nano\meter}$ which controls the typical confinement length of the edge states, or than the magnetic length $l_B=\sqrt{\frac{\hbar}{eB}}\simeq\SI{80}{\nano\meter}$ at $B=\SI{100}{\milli\tesla}$. It indicates that the edge states are broadened, for example by shallow potential fluctuations and puddles. Such very large values of $w_0$ have also been recently reported in Ref.\onlinecite{kamata2022}, and similarly attributed to charge puddles. They result in an increased capacitance $C_{\rm QH}$ accounting for the gate-puddle coupling, an increased transverse width $w_0$ and equivalently to a reduced velocity $v$, irrespective of the edge confinement depletion length $l\simeq d$.

At higher fields ($B>B_c\simeq\SI{2}{\tesla}$ in Fig.\ref{figure:QHData}a), the velocity $v$ strongly departs from this simple law, and shows strong oscillations. This crossover may be attributed to a reduced number of edge states, forming compressible and incompressible stripes under the influence of strong e-e interactions (CSG regime). We find that the crossover field $B_c$ between both regimes is approximately compatible with the heuristic law\cite{armagnat2020} $B_c\propto n^{2/3}$  (see Supplementary Material). Such oscillations have already been observed in GaAs quantum wells \cite{kumada2011} and originate from the transverse compression and decompression of plasmons when a new incompressible stripe nucleates in the bulk of the material at integer filling factors, and is progressively pushed towards the edges of the sample as $\nu$ increases (see Fig.\ref{figure:QHData}c). It is worth noting that oscillations of the velocity have also been observed in ungated graphene \cite{kumada2014} and InAs quantum wells\cite{kumada2020}, with opposite behavior (minimal widths for integer filling factors), and are then ascribed to another mechanism, namely enhanced dissipation due to a conducting bulk.

Therefore, we continue the analysis by plotting the width $w$ obtained from Eq. \ref{eq:velocity_QH} as function of the filling factor $\nu$ (see Fig.\ref{figure:QHData}b). At high filling factors $\nu\gg 15$ (i.e. low magnetic fields), $w$ slowly converges towards its saturation value $w_0$. For low filling factors, we observe that $w$ is maximum (i.e. the velocity $v$ reaches its minima) at integer filling factors. The oscillations are very strongly visible at high densities $n>\SI{3e11}{\per\square\centi\meter}$, when screening is strong and thus when the electrostatic disorder is less influential. In contrast, the oscillations are washed out at low densities. The oscillations of $v$ and $w$ are also visible though much fainter in the Pd sample (see Supplementary Material), as can be expected in a more disordered sample. 

As shown in Ref.\onlinecite{kumada2011}, the oscillations of $w$ allow for reconstructing the edge density profile. We define the local density as $x \mapsto n_e(x)= n f(x)$ ranging from $n_e(x=0)=0$ at the quantum well edge to $n_e(x)=n$ deep in the bulk of the material, as depicted in Fig.\ref{figure:QHData}c. The reconstruction is based on the following principles. The plasmon width $w$ is essentially defined by the position of the innermost edge state (compressible stripe), located at a position $x_{\rm QH}$ such that the local filling factor $\nu_e(x_{\rm QH})= \frac{hn_e(x_{\rm QH})}{eB}= \lfloor \nu \rfloor$, i.e. is the largest integer inferior or equal to the bulk filling factor $\nu$. As $\nu$ varies, $w$ spans a large range of values from $~0$ (strongly confined plasmons) to $w\simeq w_0$ (loosely confined plasmons), reflecting the variations of $x_{\rm QH}$, thus yielding an implicit equation connecting $w, B$ and $n_e$. Accounting for a broadening $w_p$ of the transverse width due to puddles, we find that the edge profile function $x \mapsto f(x)$ can be reconstructed using the implicit equation (see Supplementary Material)
\begin{equation}
f(w-w_p/2)= 1-\frac{1}{2\nu}
\end{equation}

The results are presented in Fig.\ref{figure:QHData}d. For all triplets $(n, B, w)$, we plot $1-1/2\nu$ as function of the measured width $w$. The data points describe the reconstructed edge profile, which is found to be mostly independent of the bulk density $n$. We then fit the reconstructed profile with the heuristic function $f(x)=\sqrt{\frac{x}{x+l}}$ used in Ref.\onlinecite{kumada2011} to obtain an estimate of the edge depletion length $l$. We find a good agreement with $l\simeq 60 - \SI{150}{\nano\meter}$, and a puddle broadening $w_p\simeq \SI{1.2}{\micro\meter}$ (almost identical for both Au and Pd samples, see Supplementary Material for more data sets). In particular, the depletion occurs on a scale $l$ on the order of 1 to 3 times the distance between the quantum well and the gate $d$, as anticipated from the electrostatic potential created by the gate. Besides, the different characteristic lengths $l$ and $w_0$ differ by more than one order of magnitude, while they should both be of the order of $d$ in a clean edge potential.

\begin{figure}[h!]
\centerline{\includegraphics[width=0.4\textwidth]{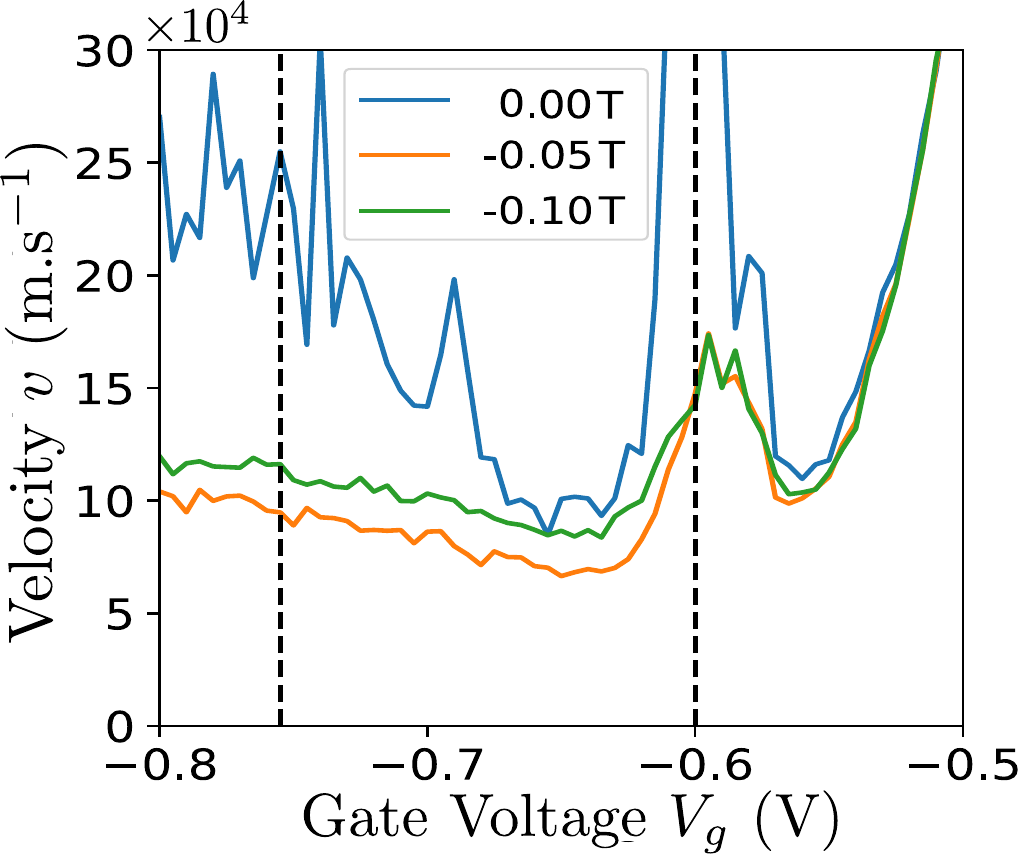}}
\caption{{\bf Velocity measured in the QSH regime, at $B\simeq0$:} Velocity $v$ as function of the gate voltage $V_g$ applied to DCg, for three values of the magnetic field $B$ close to $B=0$. The gap region estimated from the resistance $R_{2T}$ is indicated by vertical dashed lines.
}\label{fig:QSHData}
\end{figure}

\paragraph{Plasmons at zero magnetic field --}

We now analyze the measurements at $B=0$ when the gate voltage $V_g$ is tuned to adjust the Fermi level in the gap of the material. Given the insulating bulk, and the much faster response times of edge states compared to bulk states \cite{dartiailh2020}, we argue and assume in the following that the phase response is dominated by edge transport, and that the velocity is that of the QSH edge channels. The following analysis supports this assumption.

The amplitude of the signal is rather weak in this regime, and consequently the phase measurements more scattered. Nonetheless, we reliably observe (in all samples and configurations) small velocities (see Fig.\ref{fig:QSHData}), on the order of $v=v_{\rm QSH}\simeq \SI{10e4}{\meter\per\second}$ for the Au sample (\SI{2e4}{\meter\per\second} for the Pd sample). We point out that these values are significantly smaller than those predicted by the band structure \cite{krishtopenko2018,strunz2020}, which is only slightly smaller than the Fermi velocity in the conduction band $v_{\rm F}^{\rm CB}\simeq \SI{1e6}{\meter\per\second}$.

As the gate voltage $V_g$ is driven towards the valence band ($V_g\lesssim \SI{-0.75}{\volt}$), the velocity is found to increase again ($v\gtrsim\SI{2e5}{\meter\per\second}$). This is in line with the expected Fermi velocity in the valence band $v_{\rm F}^{\rm VB}\simeq \SI{2e5}{\meter\per\second}$ (though this estimation is made difficult by the camelback structure of the valence band, and its strong variations with parameters such as the quantum well thickness $t$).

\paragraph{Discussion --}

QH and QSH edge states have different origins, namely the formation of the Landau level spectrum for QH edges states vs the topological band inversion of HgTe for the QSH ones. However, their exact properties could both be affected by electrostatic disorder, and therefore may be correlated to one another. Though the following considerations are more speculative, we put forward examples of such relations.

The velocities in these two regimes can not directly be compared ($v\propto B^{-1}$ in the QH case). However, focusing first on the Au-gated sample, we observe that the two capacitances are very close to each other, with \footnote{for $\epsilon\simeq 8.8\epsilon_0$, largely dominated by the contribution of the capping CdTe layer} $C_{\rm QH}=\epsilon w_0/d\simeq \SI{1.4}{\nano\farad\per\meter}$ and $C_{\rm QSH}=4e^2/hv_{\rm QSH} \simeq \SI{1.5}{\nano\farad\per\meter}$. This value is also fully compatible with the density of state previously measured in the QSH edge state \cite{dartiailh2020}, (with a gold-gated device, and accounting for a factor 3 in the distance $d$ between the two devices).
Moreover, the following ratios between the two samples $\frac{w_0({\rm Pd})}{w_0({\rm Au})}\sim \frac{v_{\rm QSH}({\rm Au})}{v_{\rm QSH}({\rm Pd})}\sim 3.8$ further corroborates that large values of $w_0$ and slow velocities in the QSH regime both originate from the electrostatic disorder, and yield $C_{\rm QH}\simeq C_{\rm QSH}\simeq \SI{5.4}{\nano\farad\per\meter}$ in the Pd sample. We note that the characteristic puddle broadening length $w_p$ is observed to be identical in both samples, for an unclear reason. We however stress that the edge profile reconstruction relies on various crude approximations, and probes a high magnetic field regime, while $w_0$ and $v_{\rm QSH}$ are obtained at low (or zero) magnetic fields.

These simple comparisons should not be overinterpreted, especially since they connect different regimes (QH and QSH). Nonetheless, they suggest that all measured quantities reflect shallow fluctuations of the electrostatic potential yielding puddles to which the different types of edge states couple. They could play an important role in understanding the causes of scattering in the edge states.

\section{Summary and Outlook}

Puddles play a minor part in archetypical studies of the quantum Hall effect in GaAs hetero-structures thanks to larger gaps, optimized electrostatic disorder, and the natural protection of QH edge states against scattering \cite{lin2021}. However their role has been recently stressed in the quantum spin Hall effect \cite{vayrynen2014,dartiailh2020,kamata2022,shamim2022} or the quantum anomalous effect \cite{lippertz2022}, where characteristic gap scales are much smaller. 

In this context, our analysis of plasmon velocities in the classical and quantum Hall regime ($B\neq0$) and in the QSH gap ($B=0$) examines the interplay of puddles with high-frequency edge channel transport in HgTe quantum wells. It consistently points towards the picture of edge states coupled to puddles that form due to electrostatic disorder. Though the steep edge confinement takes place over a distance $l\sim d$, the quantum Hall edge states spread at low fields over a width $w_0$ of order 1-\SI{4}{\micro\meter}. In addition we find that the velocity $v_{\rm QSH}$ in the QSH edge state regime is strongly reduced compared to the anticipated Fermi velocity of the edge channels, in agreement with recent measurements of the edge density of state in similar quantum wells.

This body of works suggest that, before e-e interaction \cite{wang2017} or other mechanisms, puddles play a prominent role in the physics of topological edge states, and constitute a serious hurdle in order to investigate the topological physics of pristine edge states. We hope that the progress in the growth and lithography of existing materials, the development of new platforms with enhanced gaps \cite{deng2020} and lesser electrostatic disorder will help overcome disorder.

\begin{acknowledgements}
The authors warmly thank S. Shamim and N. Kumada for insightful discussions, and W. Beugeling for technical support with ${\bf k}\cdot{\bf p}$ simulations. 
This work has been supported by the ERC, under contract ERC-2017-StG "CASTLES" and ERC-2017-Adv "4TOPS", the DFG (SFB 1170 and Leibniz Program), Germany’s Excellence Strategy (Cluster of Excellence Matter and Light for Quantum Computing ML4Q, EXC 2004/1 - 390534769, and the W\"urzburg-Dresden Cluster of Excellence on Complexity and Topology in Quantum Matter, EXC 2147, 39085490)  and the Bavarian Ministry of Education (ENB Graduate school on ‘Topological Insulators’, and the Institute for Topological Insulators), and finally by the JST, PRESTO Grant Number JPMJPR20L2, Japan.

\section*{Data availability}
The data sets generated and/or analyzed during the current study are available from the corresponding author on reasonable request.

\section*{Author contributions}

A.G. and E.F. performed the measurements and the data analysis, under the supervision of H.K. and E.B. A.G. fabricated the samples, with help from H.K., and based on MBE layers grown by L.L. H.K. and E.B. supervised the project. All authors participated to the analysis of the results and to the writing of the manuscript.
\end{acknowledgements}

\clearpage

\begin{center}
{{\bf {\large -- Supplementary Online Material --}}}
\end{center}

\section{Microwave measurement setup}

\begin{figure*}[h!]
\centerline{\includegraphics[width=0.7\textwidth]{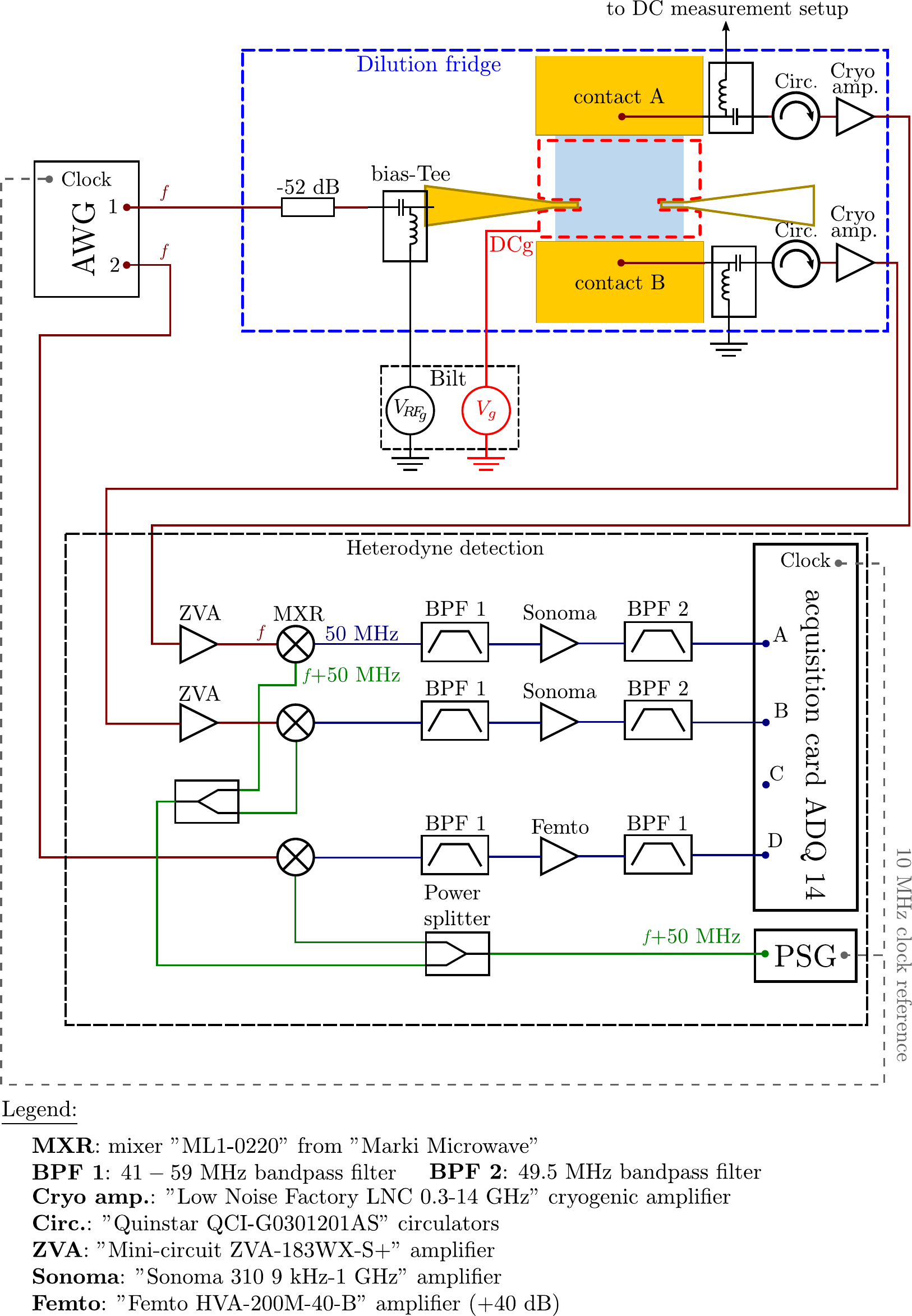}}
\caption{{\bf Sketch of the setup for RF measurements:} The setup is described in the text. The different color used here for the RF lines correspond to the different frequency of the signal: the frequency $f$ ($\sim$ GHz) in dark red, $f+\SI{50}{\mega\hertz}$ in green and $\SI{50}{\mega\hertz}$ in blue. In this sketch, we have considered an excitation from RFg 1 only, RFg 2 is turned off. The gates RFg and DCg are DC biased by voltages $V_{RFg}$ and $V_g$ generated by a iTest Bilt 2242 voltage source.}
\label{figure:RFsetup}
\end{figure*}

Our sample is placed at the mixing chamber stage of a cryogenic dilution fridge, at a base temperature of about $\SI{20}{\milli\kelvin}$. The experimental setup of the RF measurements is depicted in Fig.\ref{figure:RFsetup} and described below. A sine wave of frequency $f$ in the GHz range is generated from a Keysight M8196A arbitrary waveform generator (AWG) and sent to the finger gate RFg of the sample through the RF lines of the fridge. We set the signal amplitude to $\SI{1}{\volt}$ (maximum amplitude allowed by the AWG), and the signal is then attenuated by $\sim-\SI{52}{\decibel}$ thanks to the attenuators placed along the fridge RF lines. At the sample level, the signal has an amplitude lower than $\SI{2}{\milli\volt}$ given the finite attenuation of the RF lines (in addition to the fixed attenuators). 

After being emitted by RFg, the signal is collected by the two ohmic contacts (A and B), and then preamplified by a LNF-LNC0.3\_14B (Low Noise Factory) cryogenic amplifier with a $0.3-\SI{14}{\giga\hertz}$ bandwidth ($+\SI{37}{\decibel}$) before being sent to a heterodyne detection system at room temperature. At the output of the fridge RF lines, the signal coming from each contact is first amplified by a Mini-Circuit ZVA-183WX-S+ amplifier ($+\SI{26}{\decibel}$ gain for $\SI{700}{\mega\hertz}$ to $\SI{18}{\giga\hertz}$) before being mixed with a Local Oscillator (LO) thanks to a Marki Microwave doubled balanced mixer ML1-0220. This LO is a sine wave generated by an Agilent Technologies E8247C PSG CW signal generator (PSG) detuned from the AWG output by $\SI{50}{\mega\hertz}$. This mixing process converts the $\si{\giga\hertz}$ signal coming from the sample to a $\SI{50}{\mega\hertz}$ signal making the digital acquisition possible. This is done, after filtering and amplifying ($+\SI{32}{\decibel}$) as shown in Fig.\ref{figure:RFsetup}, by an ADQ-14 acquisition card from Teledyne SP devices. The acquisition card will demodulate the input signal by a $\SI{50}{\mega\hertz}$ sine wave and then capture the "in-phase" ($I$) and "in-quadrature" ($Q$) part of the signal.

In order to have phase-resolved measurement, a trigger signal synchronized with the AWG signal must be sent to the acquisition card. This trigger is a $\SI{50}{\mega\hertz}$ signal that is generated from the mixing between a sine wave coming from the AWG with same frequency as the signal sent to the sample and the PSG sine wave. All the instruments clocks (AWG, PSG and acquisition card) are then synchronized using the PSG $\SI{10}{\mega\hertz}$ clock signal as reference for the AWG and the acquisition card. 

Thanks to this heterodyne setup, we are able to measure in parallel signals coming from contact A and contact B in a frequency range spanning from $\SI{3.2}{\giga\hertz}$ to $\SI{12}{\giga\hertz}$. The minimum of this range ($\SI{3.2}{\giga\hertz}$) corresponds to the minimal working frequency of our detection system, in particular it is the minimal frequency of the bandwidth of the circulators (labelled "Circ." in figure Fig.\ref{figure:RFsetup}). The maximum of the frequency range ($\SI{12}{\giga\hertz}$) corresponds to the maximal frequency of the bandwidth of the mixers. In addition, no clear signal from the sample is measured at this high frequency, suggesting that it corresponds to a regime largely dominated by parasitic coupling. Overall, we have obtained the clearest data sets at $\SI{4}{\giga\hertz}$. They are shown in the main text. More data sets at different frequencies are discussed in section \ref{sec:Group}

\section{Additional data sets}

\subsection{Geometry of the Pd sample}
\begin{figure*}[h!]
\centerline{\includegraphics[width=0.6\textwidth]{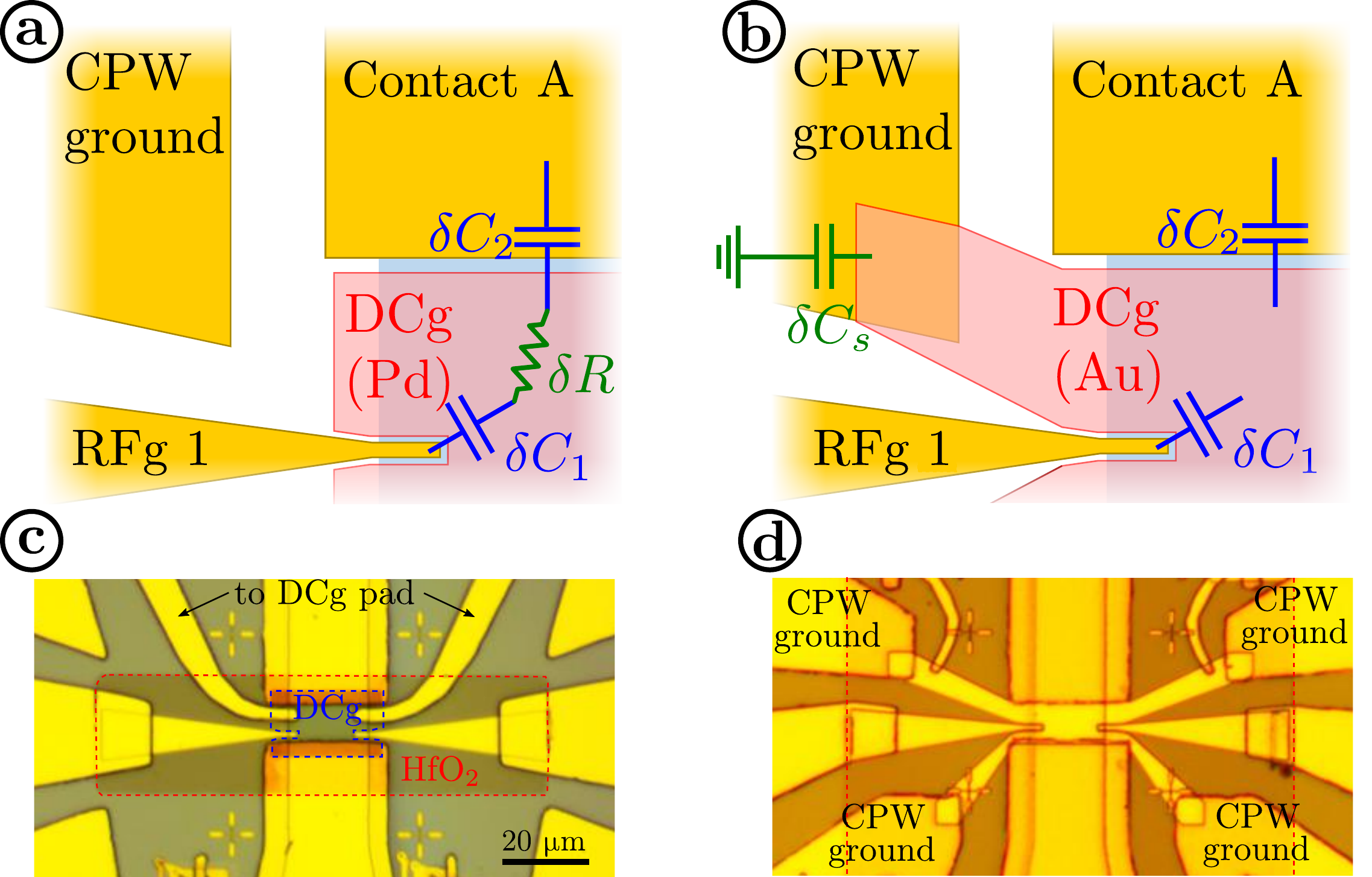}}
\caption{{\bf Comparison of the sample geometry for the Au and Pd gated samples:} a) and b) Sketch of the two different DCg designs and the parasitic coupling between RFg 1 and contact A. This parasitic coupling arises from both the coupling $\delta C_1$ between RFg 1 and the DCg and the coupling $\delta C_2$ between DCg and contact A. For DCg made of a thin layer of Pd (a), the resistivity due to the material would damp the signal passing through the DCg. This is represented in green by the resistance between $\delta C_1$ and $\delta C_2$. Using instead a conductive gate made of Au (b), the parasitic signal passing through the DCg is shunted by coupling the DCg to the CPW ground plane of the RFg. This is equivalent to placing a shunt capacitance $\delta C_s$ between $\delta C_1$ and $\delta C_2$. c) and d) Microscopic pictures of the device in the case of a Pd DCg (c) and the Au DCg (d). The region in the dashed red box corresponds to the the insulator layer made of HfO$_2$. In c) the DCg is indicated by the blue dashed box, because the thin Pd layer is difficult to distinguish.}
\label{figure:SampleGeometry}
\end{figure*}

We first briefly discuss the geometry of the Pd sample, the motivation for using a thin Pd gate, and the differences between both geometries.
Both samples are very similar, and comprise two RF gates RFg 1 and RFg 2 on both sides of the sample and which are coupled capacitively to the edge of the HgTe QW. Also, in both cases, the Fermi energy is tuned from the valence band to the conduction band thanks to a voltage bias applied at a DC gate (DCg). As seen in Fig.\ref{figure:SampleGeometry}c and \ref{figure:SampleGeometry}d, the DCg covers the whole HgTe QW mesa between the two ohmic contacts while being separated from the RFg part by a $\sim \SI{0.4}{\micro\meter}$-wide gap, in order to minimize the cross talk between the two gates. This cross-talk has been a particular concern since it contributes solely to the stray parasitic couplings between the RFg and the contact. This stray capacitance can be seen as the serial addition of the capacitive coupling $\delta C_1$ between the RFg and the DCg, and the capacitive coupling $\delta C_2$ between the DCg and the contact. In order to minimize this parasitic signal, we have tested two different geometries for the DCg sketched in Fig.\ref{figure:SampleGeometry}.

\begin{itemize}
\item The first idea is to take advantage of a resistive gate. A resistive gate would add a resistance $\delta R$ between the two stray capacitances $\delta C_1$ and $\delta C_2$ and then dissipate the RF signal passing through the DCg. In order to achieve such resistive DCg, this one will be fabricated by a thin layer of Pd. Indeed, such a material is known to have a relatively high resistivity for a metal when the thickness is low enough. We have measured that our $\sim \SI{2.5}{\nano\meter}$ thick Pd layer has a sheet resistance around $\sim \SI{300}{\ohm}/\square$ while a Au gate of $\SI{200}{\nano\meter}$ thick has a sheet resistance of $\sim \SI{7}{\ohm}/\square$. We then estimate a reduction of the parasite signal around $60\%$ compared to a pure Au gate.
\item The second idea is to ground the gate DCg in the GHz regime, by enhancing the capacitive coupling $\delta C_s$ between the DCg and the ground plane of the CPW. This is achieved by extending the DCg to overlap the 4 RFg CPW grounds over an area of $5\times\SI{5}{\square\micro\meter}$ as show in the Fig.\ref{figure:SampleGeometry}d. In this case we have estimated that $\delta C_s \sim \SI{100}{\femto\farad}$        
\end{itemize}

Both samples in fact showed very similar results, and there is in particular no sign that the Pd gate shows reduced screening. This is likely due to insufficient sheet resistance of the Pd layer, which remains far from that of ZnO layers ($ \sim \SI{1e5}{\ohm}/\square$) used in Ref.\onlinecite{kumada2020}.

\subsection{Crossover between low- and high-field regimes}

 \begin{figure*}[h!]
\centerline{\includegraphics[width=1\textwidth]{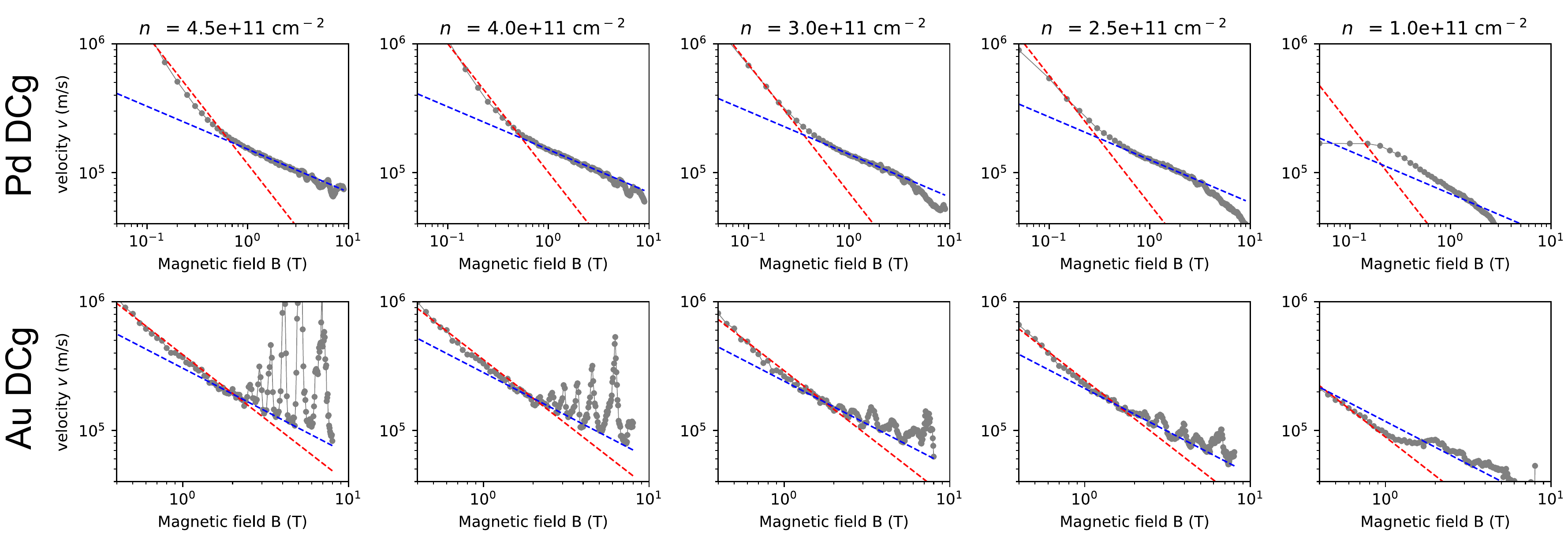}}
\caption{{\bf Phase velocity $v$ linecuts} The velocity $v$ is plotted as a function of the magnetic field $B$ (in logarithmic scales) for the Pd sample (top row) and Au sample (bottom row), for different electron densities $n$ (indicated on the top of each column). The blue dashed line corresponds to the inverse function $\alpha_1/B$ fitting. The red dashed line corresponds to the fitting law $\alpha_{1/3}/B^{1/3}$ (resp. $\alpha_{2/3}/B^{2/3}$) for Pd sample (Au sample). The black dotted lines indicate the critical magnetic field $B_c$.}
    \label{fig:FitsVvsB}
\end{figure*}

\begin{figure*}[h!]
    \centering
    \includegraphics[width = 0.7\textwidth]{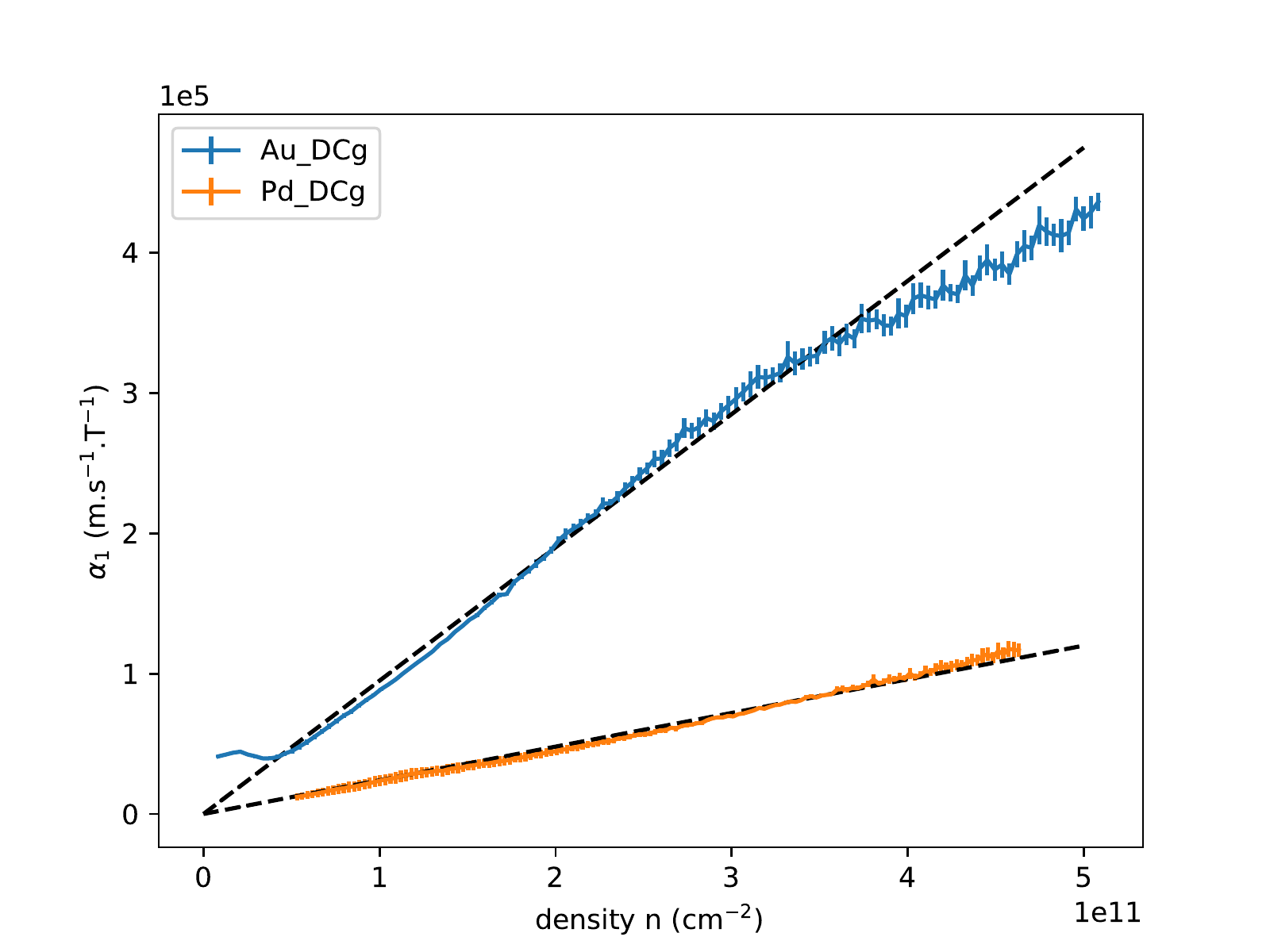}
    \caption{{\bf Determination of $w_0$:} the coefficients $\alpha_1$ from the fits of $v(B)$ are plotted as a blue solid line, as function of the carrier density $n$ for both samples. The error bars are computed from the residuals of the fits. We observe that $\alpha_1$ verifies $\alpha_1\propto n$ in a wide range of densities $n$, and extract the width $w_0$ from the slope of the linear fits shown as black dashed lines.}
    \label{fig:ExtractionWidth}
\end{figure*}  

According to Refs. \onlinecite{aleiner1994, johnson2003}, the plasmon velocity $v$ exhibits a power law $B^{-1}$ given by $v=\frac{ned}{\epsilon w B}$ if one assumes the other parameters $n,\ w$ to be constant. To confirm this, we plot $v$ as a function of $B$ on a log-log scale for different carrier densities $n$ in the conduction band, presented in Fig.\ref{fig:FitsVvsB}. These plots clearly show two power law regimes (with negative exponents) in $B$ as represented by the blue and red dashed lines: 
\begin{itemize}
    \item at low magnetic field ($B\ll B_c$), the velocity follows the expected inverse law in $B$: $v = \frac{\alpha_1}{B}$, where $\alpha_1$ is a fitting parameter.
    \item at high magnetic field ($B\gg B_c$), we observe in some parameter range a good agreement with negative power laws with different exponent ($-1/3$ for Pd DCg and $-2/3$ for Au DCg), though we don't have models describing such a behavior. Note also that for Au DCg sample, we have only considered the minimum in the oscillations of $v$.
\end{itemize}

For each sample, we plot the fitting parameters $\alpha_1$ as a function of the density $n$. This is presented in Fig.\ref{fig:ExtractionWidth}. As depicted in this figure by a black dashed line, $\alpha_1$ is linear with the density $n$ in a wide range of densities, which allows to univocally identify a constant (independent of $n$) transverse width $w_0$ in this regime (low magnetic field $B\ll B_c$). Thanks to this linear fit, one can extract this constant width $w_0$ and find $w_0 \simeq \SI{4.6}{\micro\meter}$ for Pd DCg and $w_0 \simeq \SI{1.2}{\micro\meter}$ for Au DCg.

\begin{figure*}[h!]
    \centering
    \includegraphics[width = 0.65\textwidth]{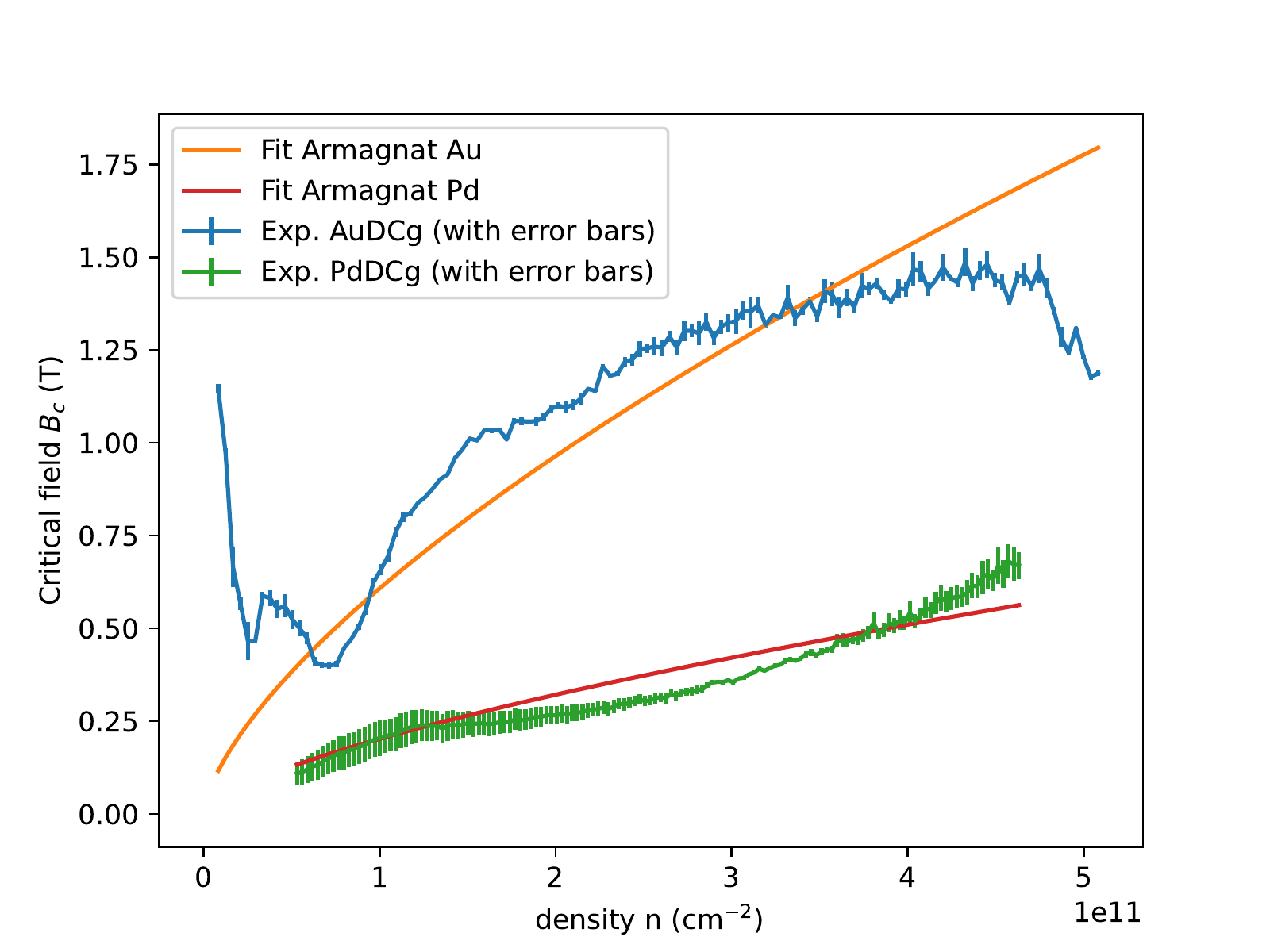}
    \caption{{\bf LB to CSG crossover: }The critical field $B_c$  in $B$ data plotted as a function of the electron density $n$ for the Pd and Au gates samples. The dots (with error bars) are the data points, while the solid lines correspond to the scaling law of the critical field $B_c\propto n^{\frac{2}{3}}$ predicted by Armagnat \textit{et al.} between the LB and CSG regime\cite{armagnat2020}.}
    \label{fig: velocities - armagnat}
\end{figure*}

One possible scenario to explain the transition between a low- and a high-field regime is the transition between a Landauer-Büttiker (LB) regime \cite{buttiker1988,halperin1982} and a Chklovskii-Shklovskii-Glazman (CSG) regime \cite{chklovskii1992} of the QH edge channels, more recently numerically revisited in Armagnat \textit{et al.} \cite{armagnat2020}. Though this is not central to our argumentation, we explore more in detail this possibility in this section.
At low magnetic field the edge channels are narrow, only spreading over a characteristic width given by the magnetic length $l_B =\sqrt{\hbar/eB}$. In this regime, the electronic interactions are neglected and the electrostatics is dominated by the transverse confining potential $U(x)$ of the QW (the coordinate $x$ describing the position transverse to the propagation direction). This LB description is not valid anymore for high magnetic field. Instead, a more appropriate description is given by the CSG picture, in which electronic interactions are now considered and play a significant role on the electrostatics of the system \cite{chklovskii1992}. The consequence is that the edge states acquire a finite width, constituting compressible stripes, separated by insulating regions called incompressible stripes. The transition between these two regimes happens when the magnetic length $l_B$ becomes comparable to one edge channel typical width $a\sim d/\nu$. According to this statement, the crossover magnetic field $B_c$ should verify $B_c\propto n^{\frac{2}{3}}$. In Fig.\ref{fig: velocities - armagnat}, we have plotted the critical field $B_c$ as a function of the bulk electron density $n$, and compare it to this power law. Though the agreement is not very good, the increase of $B_c$ with the carrier density is captured, suggesting a similar mechanism for the transition in our sample.

\subsection{Velocity oscillations in the Pd and Au samples}

\begin{figure*}[h!]
\centering
\includegraphics[width = 0.95\textwidth]{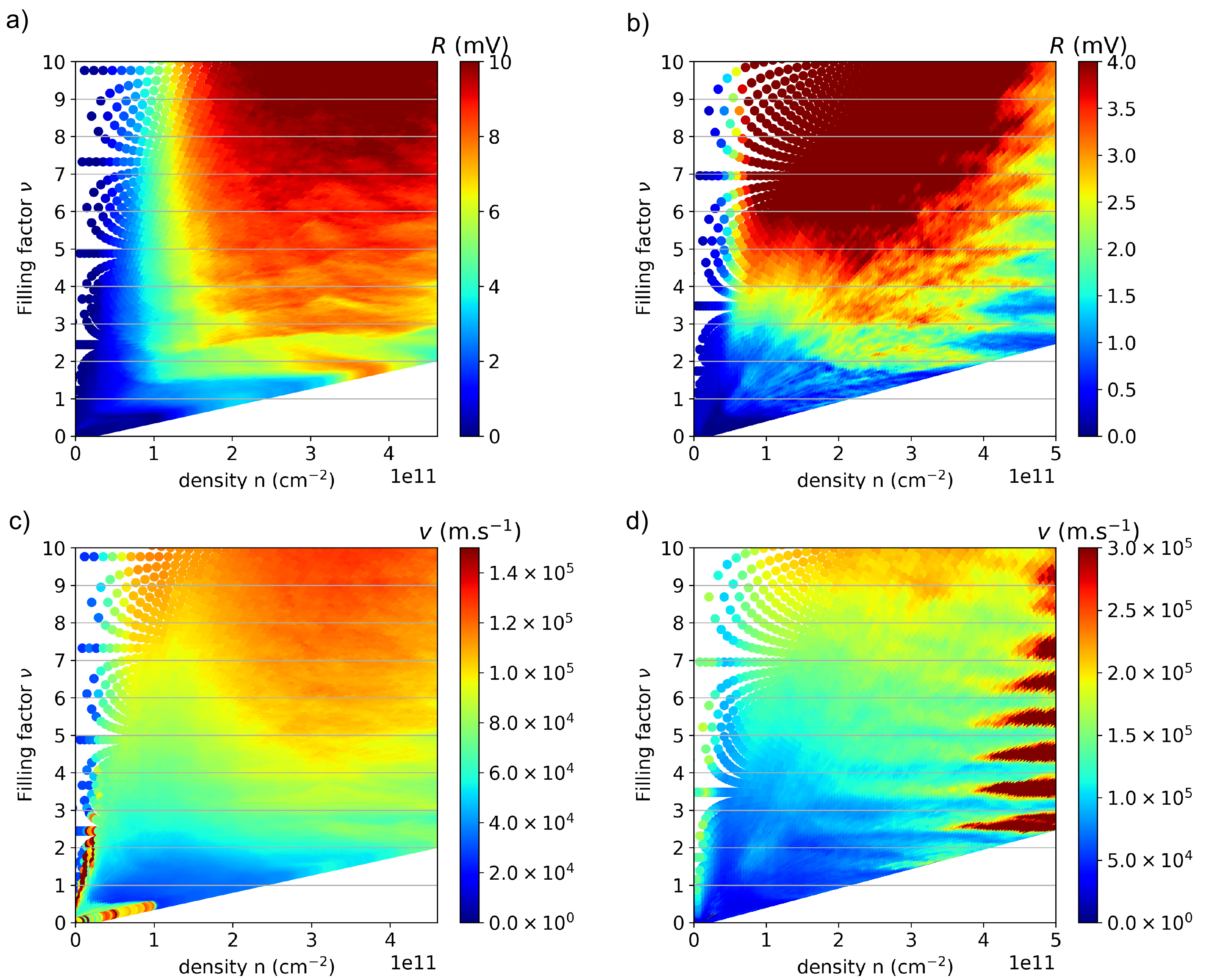}
\caption{{\bf Colorplots of the RF amplitude $M$ and velocity $v$ measured in the Au and Pd samples:} The measured magnitude $M$, as well as the velocity $v$ (computed from the phase $\phi$) are shown in a colorplot as function of the electron density $n$ and filling factor $\nu$ for both the Pd (a, c) and Au (b, d) samples. The oscillations of the velocity are clearly visible in the velocity measurements of the Au sample at high density. Both sets have been measured at a frequency $f=\SI{4}{\giga\hertz}$.}
\label{fig:QHColorplots}
\end{figure*} 

We show in this section full colormaps of the RF signal amplitude $M$ and velocity $v$ measured on both samples with Au and Pd gates. In Fig.\ref{fig:QHColorplots}, one observes that as expected both $M$ and $v$ increase when $\nu$ increases. In the high density region $n>\SI{2e+11}{\per\square\centi\meter}$, oscillations in the velocity $v$ become visible, especially for the Au-gated sample. The features align very well with the integer values of $\nu$, supporting the claims of the main text.

\subsection{Edge potential reconstruction in the Pd and Au samples}
\begin{figure*}[h!]
\centering
\includegraphics[width = 0.8\textwidth]{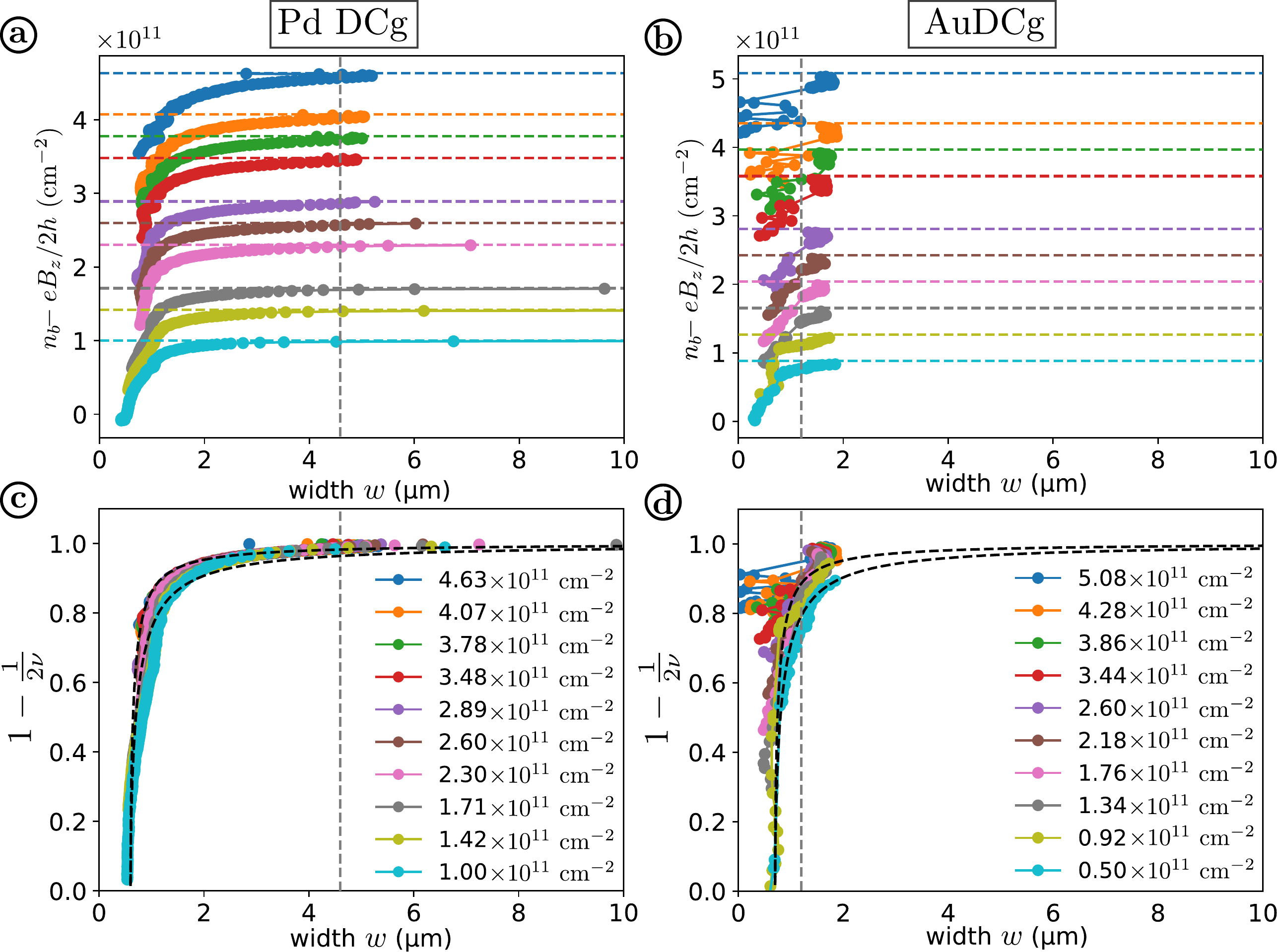}
\caption{{\bf Mapping of the edge density profile $n_e(x)$:} The edge density profile $n_e(x)$ is reconstructed by plotting the quantity $n - eB/2h$ as a function of the plasmon width $w$ (a) and b)), and the normalized quantity $1-1/2\nu$. c) and d) For both samples. the different colors correspond to different carriers density as indicated in the legend, for Pd DCg and Au DCg respectively. The black dashed curve corresponds to the fits with the test function $f(x)$ with the parameter values $l = \SI{60}{\nano\meter}$ and $l = \SI{150}{\nano\meter}$.}
\label{fig:EdgeDensityProfile}
\end{figure*} 

According to the theory developed for screened plasmons \cite{aleiner1994,johnson2003,kumada2011}, the plasmon width $w$ is fixed by the innermost incompressible strip, localized at the transverse position $x_{\rm QH}$. As stated in the main text, we assume that the edge state is further broadened by puddles, and write $w=w_p/2+x_{\rm QH}$. The plasmon width oscillates between a maximum $w\simeq w_0$ for bulk filling factor just exceeding an integer value (i.e. with an edge state nucleating in the bulk) $\nu \simeq \lfloor \nu \rfloor^+$, and a minimum $w\simeq 0$ when the bulk filling factor is slightly below the next integer number $\nu \simeq \lceil \nu \rceil^{-}$. 

The local density at the position $x_{\rm QH}$ of the innermost edge channel are actually equal at zero magnetic field and for a finite filling factor \cite{armagnat2020}, namely $n(x_{\rm QH}, B) = n(x_{\rm QH},0)=n(w-w_p/2)$. As the position of the edge state $x_{\rm QH}$ is related to the local filling factor $\nu_e(x_{\rm QH})= \frac{hn_e(x_{\rm QH})}{eB}$ being an integer, i.e. $\nu_e(x_{\rm QH}) = \lfloor \nu \rfloor$, one can connect the bulk density $n$, the local edge density $n_e(x_{\rm QH})$ and the magnetic field $B$, so that one can map $n_e(x)$ as the plasmon width $w$ varies and obtain:
\begin{equation}
n_e(x_{\rm QH}) = n_e(w-w_p/2) = \frac{e|B|}{h}\lfloor \nu\rfloor
\label{eq:EdgeDensityProfile0}
\end{equation}
At half filling $\nu = \lfloor \nu \rfloor + 1/2$, this further simplifies to:
\begin{equation}
n_e(x_{\rm QH}) = n_e(w-w_p/2) = \frac{e|B|}{h}\left(\nu - \frac{1}{2}\right) = n - \frac{e|B|}{2h}
\label{eq:EdgeDensityProfile}
\end{equation}

To account for disorder and smearing effects, we assume, that on average, Eq.(\ref{eq:EdgeDensityProfile}) can be generalized to all filling factors $\nu$, considering the approximation that the plasmon width $w$ at $\nu = \lfloor \nu \rfloor + 1/2$ is the mean value of the width for the whole filling factor interval $\left[\lfloor \nu \rfloor, \lfloor \nu \rfloor + 1\right[$ in which the number of egde states is fixed at $\lfloor \nu \rfloor$.

 In Fig.\ref{fig:EdgeDensityProfile}a and \ref{fig:EdgeDensityProfile}b we have plotted the quantity $n - \frac{e|B|}{2h}$ as a function of the plasmon width $w$ for the Pd DCg and Au DCg samples respectively and for different bulk electron densities $n$ as well as the normalized quantities $\left(n - \frac{e|B|}{2h}\right)/n = 1-1/2\nu$, where one can observe that the curves for the different densities are superimposed. This confirms that the shape of the carrier density profile $n_e(x)$, and in particular its characteristic depletion length $l$, does not depend much on the bulk carrier density $n$. For comparison, one can fit the data in Fig.\ref{fig:EdgeDensityProfile} with a heuristic edge function given by Kumada \textit{et al.} \cite{kumada2011}: 
\begin{equation}
f(x)=\frac{n_e(x)}{n} = \sqrt{\frac{x}{x+2l}}
\label{eq:KumadaFunction}
\end{equation}
Fits to this function yield a characteristic length $l\sim 60-\SI{150}{\nano\meter}$ which is on the order of $d\sim \SI{66}{\nano\meter}$, and a plasmon broadening length $w_p\simeq\SI{1.2}{\micro\meter}$ for both samples.

\subsection{Velocity in the QSH regime}

\begin{figure*}[h!]
\centering
\includegraphics[width = 0.85\textwidth]{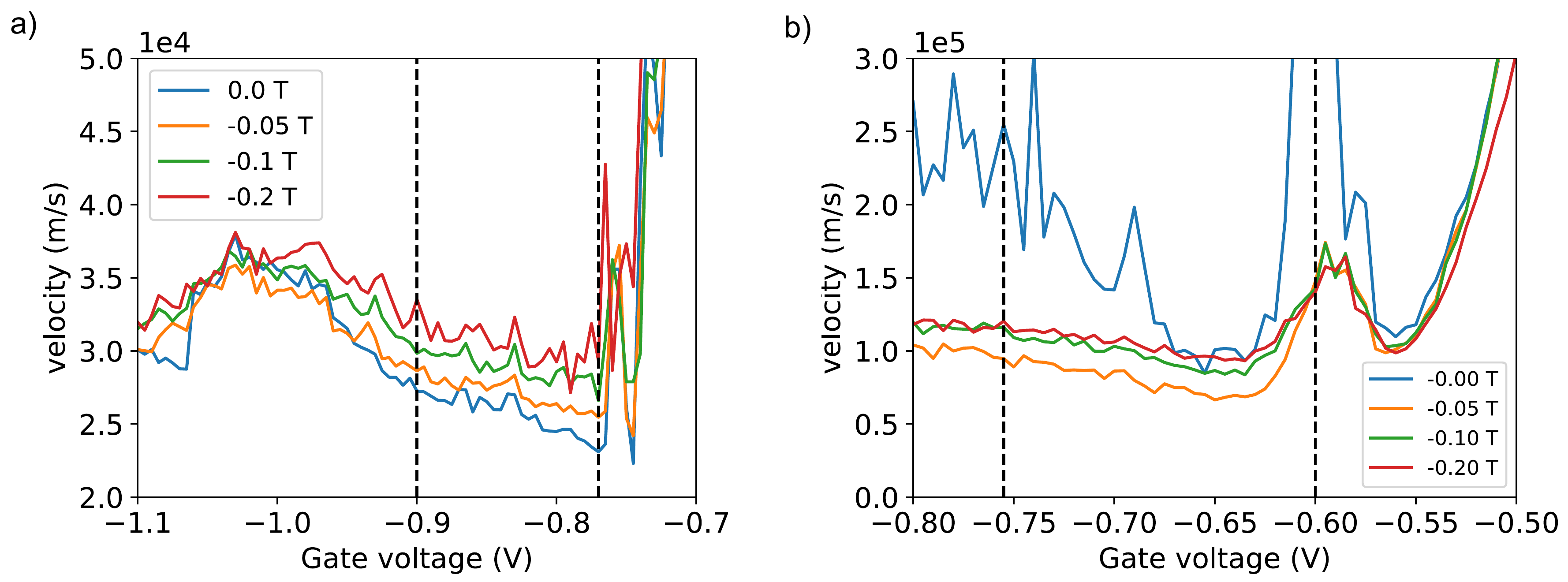}
\caption{{\bf Velocity measured in the QSH regime, at $B\simeq0$:} Velocity $v$ as function of the gate voltage $V_g$ applied to DCg, for three values of the magnetic field $B$ close to $B=0$, and for Pd (a) and Au samples (b). The gap region estimated from the resistance $R_{2T}$ is indicated by vertical dashed lines.}
\label{fig: velocities - QSH v mag linecuts}
\end{figure*}

We here focus on the velocity $v$ in the QSH regime (i.e. at $B=0$ near the gap), plotted as a function of gate voltage (Fig.\ref{fig: velocities - QSH v mag linecuts}). It exhibits a minimum in the gap, down to $\sim \SI{2.5e4}{\meter\per\second}$ for Pd DCg and $\sim \SI{1e5}{\meter\per\second}$ for Au DCg. As mentioned in the main text, these velocities are much lower than the ones predicted ${\bf k}\cdot {\bf p}$ calculations of the band structure, which predict plasmon velocities lower than but comparable to the Fermi velocity $v_F^{\rm CB}\simeq\SI{1e6}{\meter\per\second}$ in the conduction band.

\subsection{Issues in the calibration}

\begin{figure*}[h!]
\centering
\includegraphics[width = 0.5\textwidth]{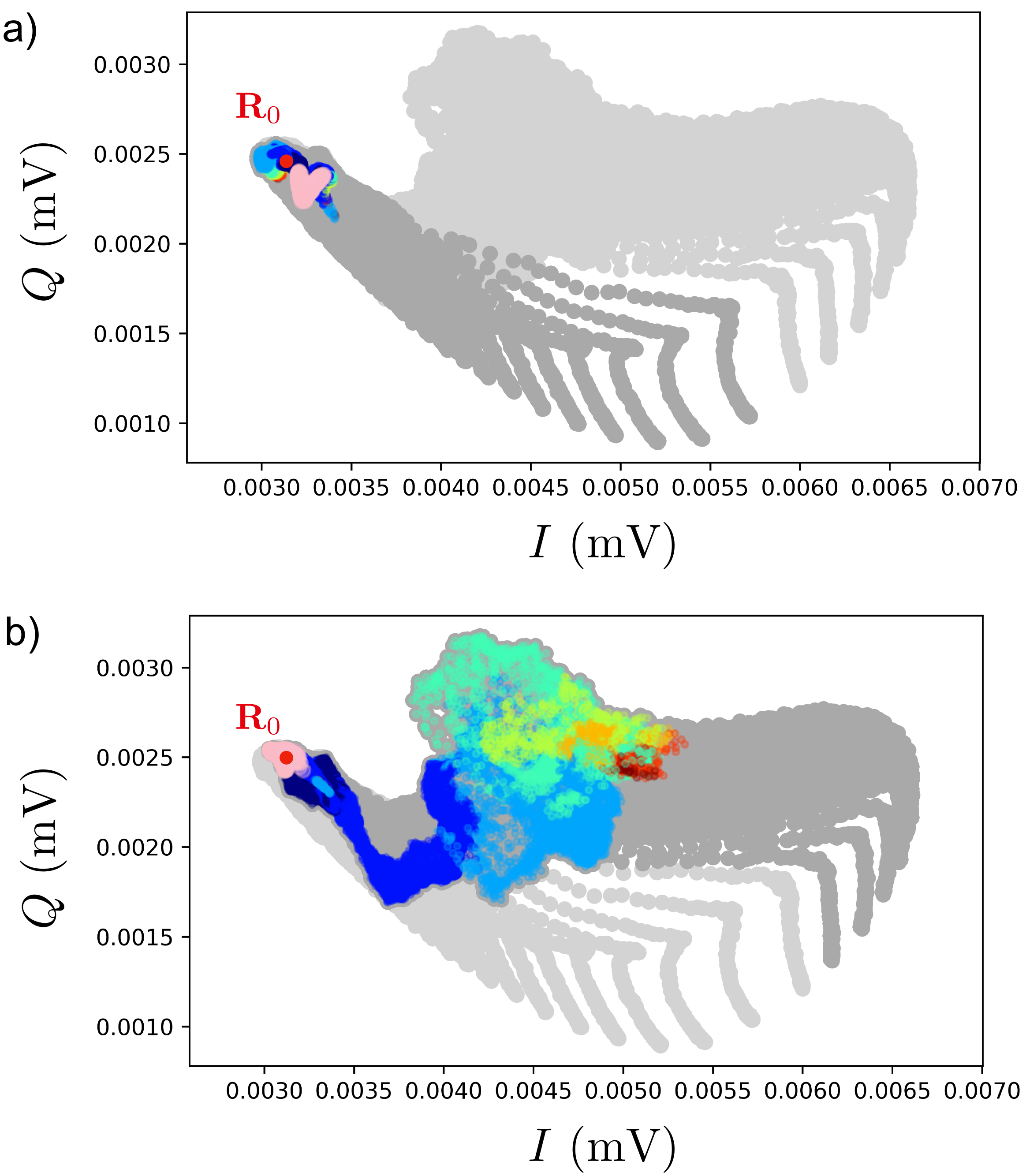}
\caption{{\bf Example of an ill-defined calibration:} As in the main text, the same data set is represented in both panels in the Nyquist complex plane ($I$,$Q$), as light or dark or colored dots. In Fig a) (resp b)), each data point is colored according to its filling factor $\nu$ measured in the DC two-terminal resistance $R_{2T}$. $B<0$ (resp $B>0$) data points are darkened for clarity. Though the parasite stray coupling vector ${\mathbf R}_0$ can be properly determined here (see dark red dot), it is impossible to define a phase reference $\phi_0$ for which the velocities do not change signs.}
\label{fig:BadCal}
\end{figure*}

In this section, we discuss problems encountered when calibrating data sets in other samples or experimental runs, as well as a plausible origin. We recall that we calibrate the data by i) subtracting a parasitic stray coupling, measured in a situation where the edge states do not contribute to the signal (reversed magnetic field or carrier polarity) ii) subtract a phase reference taken when plasmons propagate quasi-infinitely fast, here in the limit of metallic 2D plasmons (at $B=0$ and high densities $n>\SI{4e11}{\per\square\centi\meter}$). We have for examples observed that some data sets have the following issues, when we conduct the calibration procedure described in the main text:
\begin{itemize}
\item Near $B=0$, sweeping $V_g$ from positive towards negative values, the phase does not wind around the origin in a given direction (clockwise) but winds in the counter-clockwise direction in some range of $V_g$. As a consequence, the velocity $v$ shows divergences and sign changes near $B=0$. When $B$ increases, the phase winding is normal and, despite the ill-defined calibration, the results described in the main text for large $B$ can be verified in these data sets as well. 
\item Some samples show no strong chirality: the signal amplitude $M$ does not cancel out in any of the two directions of the magnetic field.
\end{itemize}
Though we do not have a precise understanding of the situation, we point out that the typical transverse widths $w_0, w_p$ of the edge states are comparable to the distance $L$ between the finger gates $RF_g$ and the contacts A and B. Assuming $w_0, w_p$ determine the size of typical puddles, disorder could result in complex percolation paths connecting RF$_g$ and the contacts A and B regardless of the chirality imposed by the magnetic field, making the calibration impossible in particularly disordered samples.

\section{Frequency dependence: study in the Pd sample}

\subsection{Calibrated data for different frequencies}

We here analyze the magnitude of the signal for different frequencies. As mentioned in the main text, the signal magnitude $M$ has a strong asymmetry in magnetic field $B$: it has a vanishing value for $B>0$ and is non zero for $B<0$. This feature is the signature of the chirality of the QH edge states. To illustrate this further, we define the asymmetry function of the magnitude $\chi(V_g, B)$ in the $(V_g, B)$ plane as:
\begin{equation}
\chi(V_g, B) = \frac{M(V_g, B) - M(V_g, -B)}{M(V_g, B) + M(V_g, -B)}.
\end{equation}

This function varies between extremal values $\pm 1$  that are reached when $M$ is strictly zero for one $B$ polarity only, and vanishes when the signal magnitude is equal for positive and negative $B$. 

\begin{figure*}[h!]
\centerline{\includegraphics[width=1.1\textwidth]{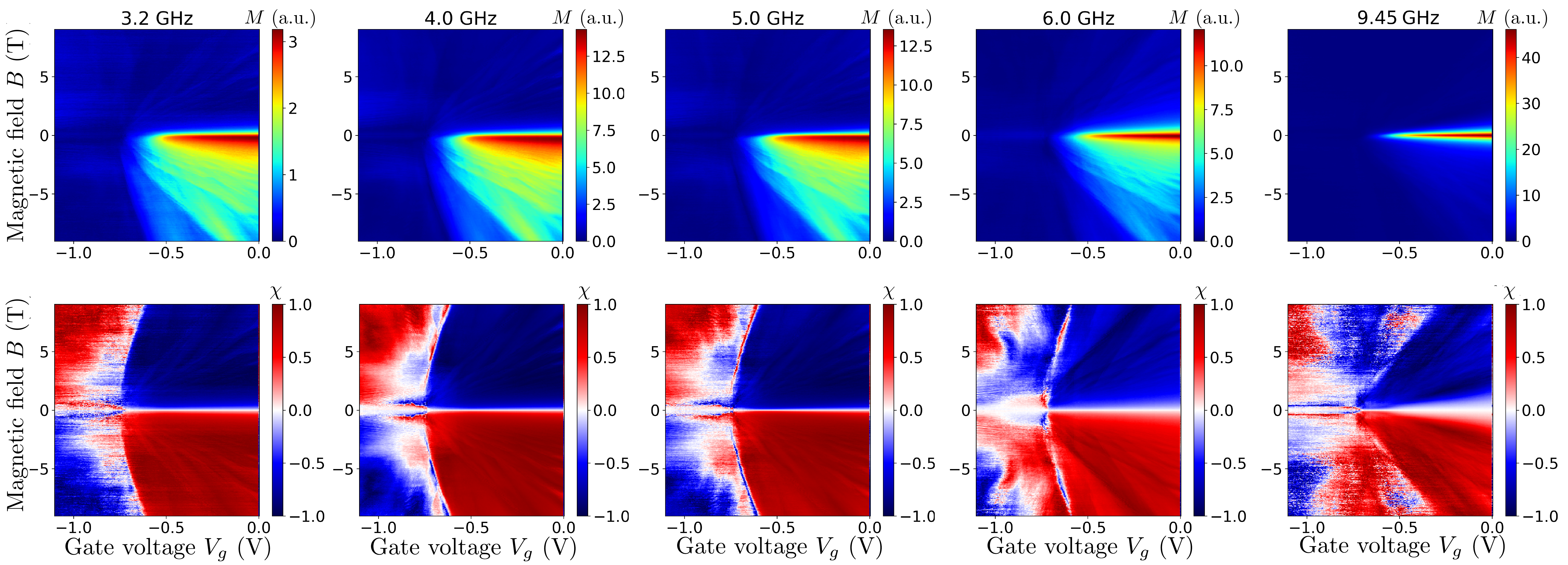}}
\caption{{\bf Measured RF magnitudes at different frequencies:} In top panels are represented the different measured RF fan charts of the signal magnitude $M$ for different frequencies (indicated on the top of each figure). In the bottom panels are represented the chirality function $\chi(V_g,B)$ of each magnitude RF fan chart.}
\label{figure:FreqDep}
\end{figure*}

In Fig.\ref{figure:FreqDep}, we present both the RF fan chart of $M$ and color plots of $\chi$ for frequencies $f=\SI{3.2}{\giga\hertz}$, $\SI{4}{\giga\hertz}$, $\SI{5}{\giga\hertz}$, $\SI{6}{\giga\hertz}$ and $\SI{9.45}{\giga\hertz}$. Though the maximum amplitude of $M$ is strongly frequency-dependent, the colorplots look relatively similar. One can nevertheless clearly observe that the amplitude measured in the QH regime (relative to the peak value at $B=0$) decreases with frequencies, and that the asymmetry between $B$ and $-B$ weakens.  Now looking at the colorplots of $\chi$, we observe that the signal is strongly asymmetric ($\chi(V_g, B) \simeq \pm1$ for $B \neq 0$) for each frequency, in particular in the conduction band ($V_g\gtrsim -\SI{0.6}{\volt}$). For lowest frequencies $f=\SI{3.2}{\giga\hertz} \text{ to }\SI{5}{\giga\hertz}$, the asymmetry function value varies abruptly from $\chi \simeq +1$ to $\chi \simeq -1$ when sweeping the magnetic field from negative $B<0$ to positive $B>0$ value. In the valence band, and more particularly in the gap, the asymmetry function shows a lesser chirality than in the conduction band. Nevertheless, one also can see that the asymmetry function switches sign when passing from conduction band ($V_g\gtrsim -\SI{0.6}{\volt}$) to the valence band  ($V_g\lesssim-\SI{0.9}{\volt}$), indicating an inversion of the chirality. For higher frequencies $f=\SI{6}{\giga\hertz} \text{ and }\SI{9.45}{\giga\hertz}$, even if $\chi$ shows a strong asymmetry, the transition at $B=0$ is smoother and the asymmetry function takes lower extremal absolute value ($|\chi|\lesssim 1$) than for lower frequencies. This suggests that increasing further the frequency degrades the observed chirality in RF. This might be explained by cross-talk effects between opposite edges of the sample for which QH states have an opposite direction\cite{tu2018}. 

\subsection{Phase vs group velocity}
\label{sec:Group}

\begin{figure*}[h!]
\centerline{\includegraphics[width=.9\textwidth]{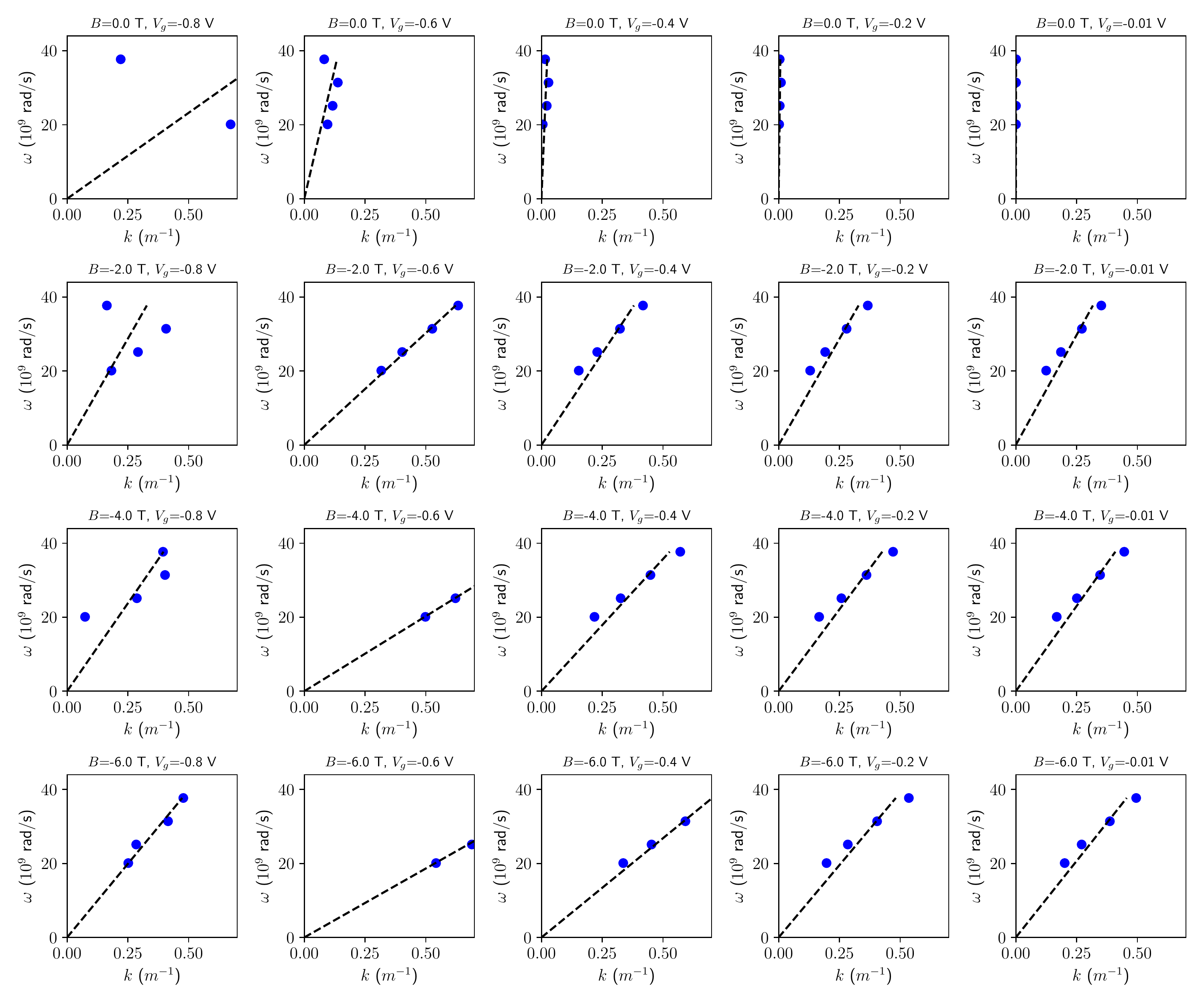}}
\caption{{\bf Dispersion relations:} The pulsation $\omega$ is plotted as function of the wavenumber $k$ computed from the measured phase shift, for various values of the magnetic field $B$ and gate voltage $V_g$ (in the conduction band). The measured data points are plotted as blue dots and corresponds to the 4 frequencies 3.2 GHz, 4 GHz, 5 GHz and 6 GHz. The dashed line is the linear fit of the data points, where the slope directly yields the group velocity $v_{g}$.}
\label{figure:LinFits}
\end{figure*}

\begin{figure*}[h!]
\centerline{\includegraphics[width=.95\textwidth]{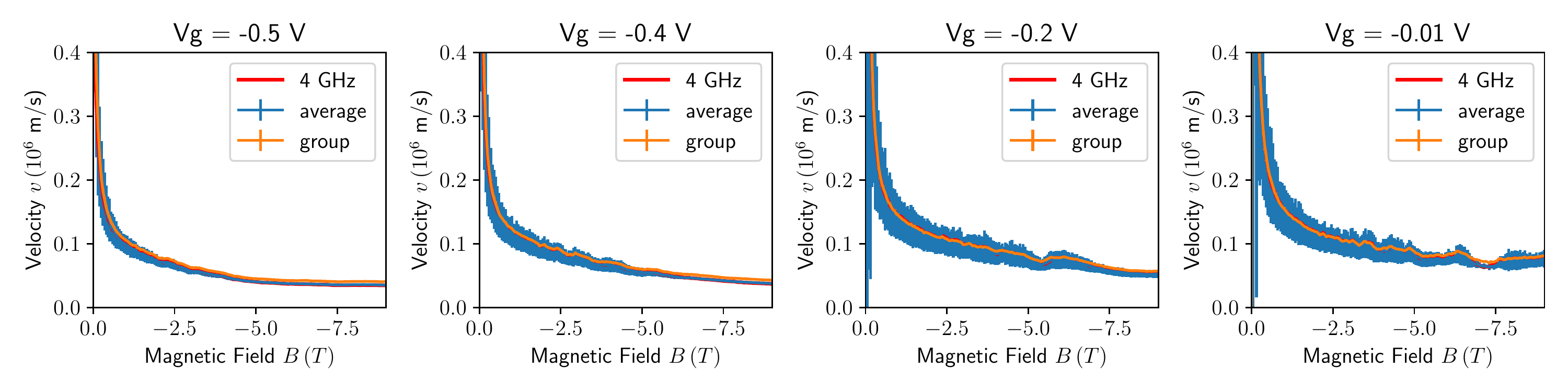}}
\caption{{\bf Comparison between the phase and group velocity}. For different values of the gate voltage indicated on the top of each figure, the phase velocity measured at 4 GHz (red solid line) is plotted as well as the extracted group velocity (blue solid line) as a function of the magnetic field $B$, with error bars estimated by the difference between the lowest and highest measured in the frequency set {3.2 GHz, 4 GHz, 5 GHz, 6 GHz, 9.45 GHz}.}
\label{figure:CompGroupPhase}
\end{figure*}

The phase velocity is defined as the velocity of a constant phase plane and it is expressed as the ratio between the wave pulsation $\omega = 2\pi f$ and its wavenumber $k$, $v_\phi = v = \frac{\omega}{k}$, while the group velocity at which energy is transported is given by $v_{g} = \frac{\partial \omega}{\partial k}$. We expect, in the studied limit of small frequencies, that the dispersion relation $\omega(k)$ is linear, such that group and phase velocity are equal, $v_\phi = v_{g}$, and the edge plasmons are non-dispersive. In the Pd sample, we have measured the phase shift fan chart for four different frequencies $3.2$ GHz, $4$ GHz, $5$ GHz and $6$ GHz. For each point in the $(V_g, B)$ phase space, we have fitted the relation between the pulsation $\omega = 2\pi f$ and the wavenumber $k = \phi/L$ with a linear function. Examples are given in Fig.\ref{figure:LinFits}. The group velocity is then extracted by measuring the slope of the linear fit. One can see that, in the conduction band, the four points are relatively well fitted with a linear dispersion (represented as a black dashed line on the figure).
Finally, in Fig.\ref{figure:CompGroupPhase}, the group and phase velocities are plotted as a function of the magnetic field $B$ for fixed values of the gate voltage $V_g$, in the conduction band. The two velocities are found to match well, confirming that one can assume that the equation $v_\phi=v_g$ is verified in our measurements.

\bibliography{Plasmons_bib.bib}

\end{document}